\documentclass[sn-standardnature]{sn-jnl}% Standard Nature Portfolio Reference Style
\usepackage{graphicx}% Include figure files
\usepackage{dcolumn}% Align table columns on decimal point
\usepackage{bm}% bold math
\usepackage{amssymb}
\usepackage{amsmath}
\usepackage{psfrag}
\usepackage{mathtools}
\usepackage{color}
\usepackage{xcolor}
\usepackage[export]{adjustbox}
\usepackage[english]{babel}
\usepackage[T1]{fontenc}
\usepackage{stmaryrd}
\definecolor{blue1}{cmyk}{0.88,0.77,0,0}
\definecolor{orangex}{cmyk}{0, 0.37,0.97,0}
\definecolor{orangex1}{cmyk}{0, 0.13,0.39,0}
\definecolor{purple1}{cmyk}{0.17,0.53,0,0}
\definecolor{green1}{cmyk}{0.5,0,1,0}
\definecolor{blue2}{cmyk}{1,1,0,0}
\definecolor{orange2}{cmyk}{0,0.5,1,0}
\definecolor{red2}{cmyk}{0,1,1,0}
\definecolor{purple2}{cmyk}{0.35,1,0.35,0.1}
\definecolor{redweak}{cmyk}{0,0.31,0.15,0}
\definecolor{blueweak}{cmyk}{0.23,0.21,0.01,0}
\definecolor{blue2}{cmyk}{0.99,0.94,0.05,0.01}

\definecolor{f1green}{cmyk}{0.76,0.03,0.56,0}
\definecolor{f1red2}{cmyk}{0.01,0.92,0.95,0}
\definecolor{f1red}{cmyk}{0.89,0.29,0.99,0.19}
\definecolor{f1purp2}{cmyk}{0.77,0.99,0.03,0}
\definecolor{f1purp}{cmyk}{0.65,0.58,0.57,0.37}

\definecolor{f1green2}{cmyk}{0.89,0.29,0.99,0.19}
\definecolor{f3blue}{cmyk}{0.88,0.77,0,0}

\definecolor{f2blue}{cmyk}{0.3,0,0.08,0}

\definecolor{SIblue}{cmyk}{0.7,0.15,0,0}
\definecolor{SIorange}{cmyk}{0,0.8,0.92,0}
\definecolor{SIbrown}{cmyk}{0.23,0.39,0.63,0.01}
\definecolor{SIpurple}{cmyk}{0.59,0.9,0,0}
\definecolor{orangex1}{cmyk}{0, 0.13,0.39,0}
\definecolor{blueweak}{cmyk}{0.23,0.21,0.01,0}
\definecolor{purple2}{cmyk}{0.35,1,0.35,0.1}
\definecolor{redweak}{cmyk}{0,0.31,0.15,0}
\definecolor{armorange}{cmyk}{0,0.44,1,0}
\definecolor{swirlpurp}{cmyk}{0.6,0.9,0,0}

\definecolor{Vblue}{cmyk}{0.85,0.5,0,0}
\definecolor{Ured}{cmyk}{0,1,1,0}
\definecolor{iblue}{cmyk}{0.8,0.58,0,0}

\newcommand{\cc}[1]{{#1}} %PRA corrections
\newcommand{\qq}[1]{{#1}} %Tiemo corrections

\jyear{2023}%

\theoremstyle{thmstyleone}%
\theoremstyle{thmstyletwo}%

\theoremstyle{thmstylethree}%

\begin{document}

\title[Synchronization-based lossless non-reciprocal scattering]{Synchronization-based lossless non-reciprocal scattering}


\author[1]{\fnm{Tiemo} \sur{Pedergnana}}

\author[1]{\fnm{Abel} \sur{Faure-Beaulieu}}

\author*[2]{\fnm{Romain} \sur{Fleury}}\email{romain.fleury@epfl.ch}

\author*[1]{\fnm{Nicolas} \sur{Noiray}}\email{noirayn@ethz.ch}

\affil[1]{\orgdiv{Department of Mechanical and Process Engineering}, \orgname{ETH Zürich}, \state{Zürich}, \country{Switzerland}}

\affil[2]{\orgdiv{Institute of Electrical and Micro Engineering}, \orgname{EPFL}, \state{Vaud}, \country{Switzerland}}


\abstract{\qq{Breaking the reciprocity of wave propagation is a problem of fundamental interest, and a mucht-sought functionality in  practical applications, both in photonics and phononics. Although it has been achieved using resonant linear scattering from cavities with broken time-reversal symmetry, such realizations have remained inescapably plagued by inherent passivity constraints, which make absorption losses unavoidable, leading to stringent limitations in transmitted power.} In this work, we solve this problem by converting the cavity resonance into a limit cycle, exploiting the uncharted interplay between non-linearity, gain, and non-reciprocity. \qq{ Remarkably, strong enough incident waves can synchronize with these self-sustained oscillations and use their energy for amplification. We theoretically and experimentally demonstrate that this mechanism can simultaneously enhance non-reciprocity and compensate absorption.} Real-world acoustic scattering experiments allow us to observe perfect non-reciprocal transmission of audible sound in a synchronisation-based 3-port circulator with full immunity against losses.}




\maketitle

\section{Introduction}\label{Section 1}
Wave propagation is said to be reciprocal if it is symmetric with respect to exchanging the locations of any sender-receiver pair. Due to its close links to time-reversal symmetry, reciprocity is ubiquitous in physics \cite{Onsager1931_I,Onsager1931_II,Casimir1945343}. Breaking reciprocity is necessary in applications which aim to guide waves in one direction while isolating another, \qq{for example to protect a source from unwanted reflections, or to induce strong topological protection against obstacles or imperfections} \cite{Zhang2021293,Zhang2023SCIADV}. Such non-reciprocal transmission is of high practical interest and has recently been reviewed for electromagnetic \cite{Verhagen2017922,Caloz2018} and mechanical \cite{Nassar2020667,Rasmussen2021} waves. 

\begin{figure}[t]%
\centering
\begin{psfrags}
\psfrag{r}{\hspace{0cm}\textbf{\text{\tiny{\fontfamily{phv}\selectfont
\begin{tabular}{ c }
\text{\footnotesize{resonance}}  \\ 
\text{\footnotesize{reciprocal}}   
\end{tabular} }}}}
\psfrag{s}{\hspace{-1.23cm}\textbf{\text{\tiny{\fontfamily{phv}\selectfont
\begin{tabular}{ c }
\text{\footnotesize{no external sources}}  \\ 
\text{\footnotesize{(symmetry intact)}}   
\end{tabular} }}}}
\psfrag{t}{\hspace{-1.6cm}\textbf{\text{\tiny{\fontfamily{phv}\selectfont
\begin{tabular}{ c }
\text{\footnotesize{{non-reciprocal scattering}}}  \\ 
\text{\footnotesize{{(symmetry {broken})}}}   
\end{tabular} }}}}
\psfrag{u}{\hspace{-0.12cm}\textbf{\text{\tiny{\fontfamily{phv}\selectfont
\begin{tabular}{ c }
\text{\footnotesize{cavity}}  \\ 
\text{\footnotesize{resonance}}   
\end{tabular} }}}}
\psfrag{v}{\hspace{-0.12cm}\textbf{\text{\tiny{\fontfamily{phv}\selectfont
\begin{tabular}{ c }
\text{\footnotesize{cavity}}  \\ 
\text{\footnotesize{limit cycle}}   
\end{tabular} }}}}
  \psfrag{a}{\textbf{\text{\tiny{\fontfamily{phv}\selectfont
\hspace{0.0cm}a} }}}
  \psfrag{b}{\textbf{\text{\tiny{\fontfamily{phv}\selectfont
\hspace{0.0cm}a} }}}
  \psfrag{c}{\textbf{\text{\tiny{\fontfamily{phv}\selectfont
\hspace{0.0cm}b} }}}
  \psfrag{d}{\textbf{\text{\tiny{\fontfamily{phv}\selectfont
\hspace{0.0cm}c} }}}
  \psfrag{e}{\textbf{\text{\tiny{\fontfamily{phv}\selectfont
\hspace{0.0cm}c} }}}
  \psfrag{f}{\textbf{\text{\tiny{\fontfamily{phv}\selectfont
\hspace{0.0cm}d} }}}
  \psfrag{g}{\textbf{\text{\tiny{\fontfamily{phv}\selectfont
\hspace{0.0cm}g} }}}
  \psfrag{h}{\textbf{\text{\tiny{\fontfamily{phv}\selectfont
\hspace{0.0cm}h} }}}
  \psfrag{i}{\textbf{\text{\tiny{\fontfamily{phv}\selectfont
\hspace{0.0cm}i} }}}
  \psfrag{j}{\textbf{\text{\tiny{\fontfamily{phv}\selectfont
\hspace{0.0cm}j} }}}
  \psfrag{k}{\textbf{\text{\tiny{\fontfamily{phv}\selectfont
\hspace{0.0cm}e} }}}
  \psfrag{l}{\textbf{\text{\tiny{\fontfamily{phv}\selectfont
\hspace{0.0cm}e} }}}
  \psfrag{m}{\textbf{\text{\tiny{\fontfamily{phv}\selectfont
\hspace{0.0cm}f} }}}
  \psfrag{K}{\textbf{\text{\tiny{\fontfamily{phv}\selectfont
\hspace{-0.3cm}\textcolor{f1red2}{Zeeman}} }}}
  \psfrag{I}{\textbf{\text{\tiny{\fontfamily{phv}\selectfont
\hspace{-0.05cm} 0} }}}
 \psfrag{J}{\textbf{\text{\tiny{\fontfamily{phv}\selectfont
\hspace{-0.05cm} time} }}}
\psfrag{x}{\hspace{-0.21cm}\textbf{\text{\footnotesize{\fontfamily{phv}\selectfont
\begin{tabular}{ c }
\text{\tiny{\textcolor{f1purp}{radiation losses}}}  \\ 
\text{\tiny{\textcolor{f1purp}{from all ports}}}   
\end{tabular} }}}}
\psfrag{y}{\hspace{-0.28cm}\textbf{\text{\footnotesize{\fontfamily{phv}\selectfont
\begin{tabular}{ c }
\text{\tiny{\textcolor{f1purp}{lossy non-reci-}}}  \\ 
\text{\tiny{\textcolor{f1purp}{procal scattering}}} 
\end{tabular} }}}}
\psfrag{A}{\hspace{-0.36cm}\textbf{\text{\footnotesize{\fontfamily{phv}\selectfont
\begin{tabular}{ c }
\text{\tiny{\textcolor{f1purp2}{self-sustained radia- }}}  \\ 
\text{\tiny{\textcolor{f1purp2}{tion from all ports}}}   
\end{tabular} }}}}
\psfrag{z}{\hspace{-0.3cm}\textbf{\text{\footnotesize{\fontfamily{phv}\selectfont
\begin{tabular}{ c }
\text{\tiny{\textcolor{f1purp2}{lossless non-reci-}}}  \\ 
\text{\tiny{\textcolor{f1purp2}{procal scattering}}}   
\end{tabular} }}}}
 \psfrag{1}{\textbf{\text{\tiny{\fontfamily{phv}\selectfont
\hspace{-0.1cm} $1$} }}}
 \psfrag{0}{\textbf{\text{\tiny{\fontfamily{phv}\selectfont
\hspace{-0.1cm} $0$} }}}
 \psfrag{2}{{\text{\tiny{\fontfamily{phv}\selectfont
\hspace{-0.2cm} \textcolor{f1green}{$T_{1\shortrightarrow 2}$ and}} }}}
 \psfrag{L}{\textbf{\text{\tiny{\fontfamily{phv}\selectfont
\hspace{-0cm} \textcolor{f1green}{$T_{1\shortrightarrow 3}$}} }}}
 \psfrag{M}{{\text{\tiny{\fontfamily{phv}\selectfont
\hspace{-0.2cm} \textcolor{f1green}{$T_{1\shortrightarrow 3}$}} }}}
 \psfrag{N}{\textbf{\text{\tiny{\fontfamily{phv}\selectfont
\hspace{-0cm} $T_{1\shortrightarrow 2}$} }}}
 \psfrag{w}{\textbf{\text{\tiny{\fontfamily{phv}\selectfont
\hspace{-0.05cm} $f$ } }}}
 \psfrag{n}{\textbf{\text{\tiny{\fontfamily{phv}\selectfont
\hspace{-0.04cm} 1} }}}
 \psfrag{o}{\textbf{\text{\tiny{\fontfamily{phv}\selectfont
\hspace{-0.04cm} 2} }}}
 \psfrag{p}{\textbf{\text{\tiny{\fontfamily{phv}\selectfont
\hspace{-0.04cm} 3} }}}
 \psfrag{q}{\textbf{\text{\tiny{\fontfamily{phv}\selectfont
\hspace{-0.04cm} 4} }}}
 \psfrag{3}{{\text{\tiny{\fontfamily{phv}\selectfont
\hspace{-0.1cm} min.} }}}
 \psfrag{4}{{\text{\tiny{\fontfamily{phv}\selectfont
\hspace{-0.05cm} 0} }}}
 \psfrag{5}{{\text{\tiny{\fontfamily{phv}\selectfont
\hspace{-0.05cm} max.} }}}
 \psfrag{6}{{\text{\tiny{\fontfamily{phv}\selectfont
\hspace{-0.1cm} Amplitude} }}}
 \psfrag{F}{{\text{\tiny{\fontfamily{phv}\selectfont
\hspace{-0.15cm} \textcolor{f1purp2}{Port 1}} }}}
 \psfrag{G}{{\text{\tiny{\fontfamily{phv}\selectfont
\hspace{-0.25cm} \textcolor{f1purp2}{Port 2}} }}}
 \psfrag{H}{{\text{\tiny{\fontfamily{phv}\selectfont
\hspace{-0.32cm} \textcolor{f1purp2}{Port 3}} }}}
 \psfrag{C}{{\text{\tiny{\fontfamily{phv}\selectfont
\hspace{-0.15cm} \textcolor{f1purp}{Port 1}} }}}
 \psfrag{D}{{\text{\tiny{\fontfamily{phv}\selectfont
\hspace{-0.25cm} \textcolor{f1purp}{Port 2}} }}}
 \psfrag{E}{{\text{\tiny{\fontfamily{phv}\selectfont
\hspace{-0.32cm} \textcolor{f1purp}{Port 3}} }}}
 \psfrag{9}{\textbf{\text{\tiny{\fontfamily{phv}\selectfont
\hspace{-1cm} \textcolor{f1green2}{output in sync}} }}}
 \psfrag{Z}{{\text{\tiny{\fontfamily{phv}\selectfont
\hspace{-0.1cm} or} }}}
 \psfrag{U}{}
 \psfrag{:}{\textbf{\text{\tiny{\fontfamily{phv}\selectfont
\textcolor{iblue}{input}} }}}
 \psfrag{>}{\textbf{\text{\tiny{\fontfamily{phv}\selectfont
\hspace{-1.5cm} \textcolor{f1green2}{decaying disturbance}} }}}

\includegraphics[width=1\textwidth]{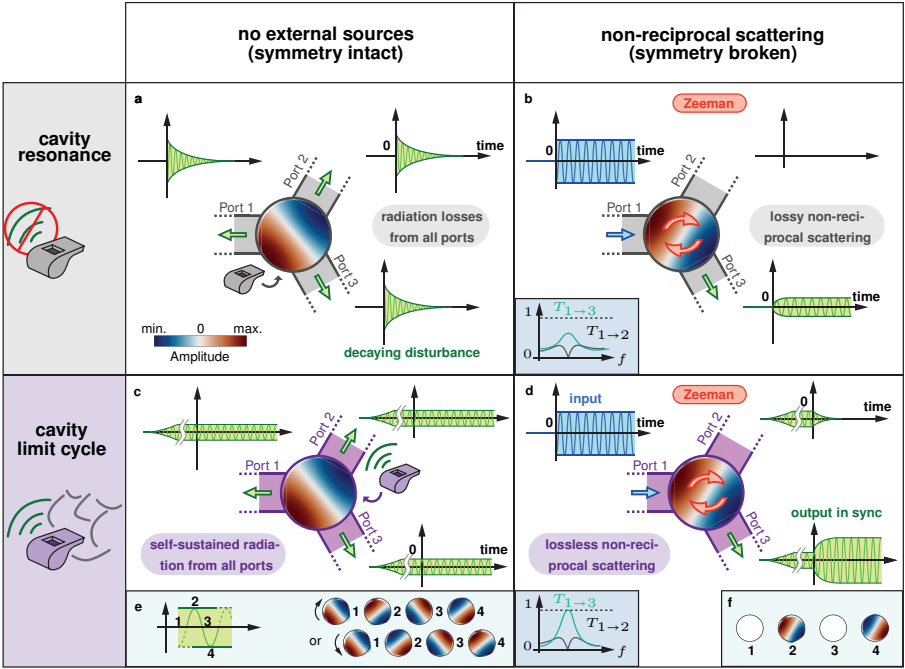}
\end{psfrags}
\caption{\textbf{Concept of synchronization-based lossless non-reciprocal scattering in a circulator}. \textbf{a}, \textbf{b} State of the art. Inducing a Zeeman-type bias in a resonator cavity (\textbf{a}) leads to non-reciprocal transmission of \qq{incident} waves (\textbf{b}). Due to the passivity of the resonance, non-reciprocal transmission will be lossy, even if the isolation of the blocked direction is perfect. \textbf{c} A limit cycle in a round cavity with multiple ports continuously radiates a signal through the ports. \textbf{d} A harmonic wave incident on the cavity can synchronize with and gain energy from the limit cycle's emissions. In the presence of a bias, this enables lossless non-reciprocal transmission.  \textbf{e} In absence of external forcing, the acoustic field is either spinning right or left.  \textbf{f} Under external forcing, the spinning acoustic field becomes a standing wave. }\label{Figure 1}
\end{figure}

Analogous to how light can be non-reciprocally routed using biased magneto-optical media \cite{Wang2005369}, unidirectional acoustic waves can be induced in a resonant cavity with a biased air flow \cite{Fleury2014516}. Both aforementioned effects are based on scattering waves at a pair of isolated, degenerate cavity modes whose symmetry is broken by a frequency splitting similar to the electronic Zeeman effect \cite{cohen2009quantum}, where an external magnetic field lifts the degeneracy of the atom's eigenstates. \qq{The exploitation of this effect to construct a circulator is well-known, and illustrated in Figs. \ref{Figure 1}a and \ref{Figure 1}b, which represents the state of the art in linear, resonance-based non-reciprocal scattering.} Of course, any macroscopic realizations, albeit inspired from atomic-scale quantum phenomena, will suffer from dissipative effects \cite{Callen195134,Kubo1966255}. The present study proposes a solution to remedy this fundamental problem by unveiling a synchronization-based mechanism for achieving lossless non-reciprocal transmission. Our method requires no internal active elements to control the field \cite{Dinc2017,Zhang2023}. 

Instead, we show here that a lossless circulator can be designed based on the non-linear \cc{interference between a wave and} a cavity limit cycle\qq{, which emits self-sustained radiation in the absence of an external source (Fig. \ref{Figure 1}c)}. \qq{The concept proposed in this work is sketched in Fig. \ref{Figure 1}d. Not only can a wave impinging on a cavity limit cycle extract energy from its radiated signal by synchronization, but the non-linear interference simultaneously \textit{enhances} non-reciprocal isolation between the ports.} In a three-port circulator with a properly tuned bias, the above scenario can enable lossless non-reciprocal transmission\qq{, overcoming the dissipation induced by low $Q$ factors in resonant cavities.} After extending classical coupled-mode theory \qq{\cite{Fan2003569,Suh20041511} to the non-linear case}, \qq{we can explain and predict} the synchronization-based transmission using only a few parameters. To validate our theory of \qq{synchronization-based lossless non-reciprocity}, we finally present an experimental realization of a loss-immune acoustic circulator, which exploits flow-induced whistling \cite{Howe1980407,Elder1982532,Bourquard2020} to create spinning acoustic limit cycles \cite{Abel2023_1,Abel2023_2}. 

%Prior to this study, a separate, in-depth experimental, numerical and theoretical investigation on the aeroacoustic instability \cc{(whistling)} in \cc{this} cavity was performed to understand its underlying physics \cite{Abel2023_1,Abel2023_2}. These works focused on the same modes (the first pair of degenerate azimuthal eigenmodes) that were used for non-reciprocal optical and acoustic transmission in \cite{Wang2005369} and \cite{Fleury2014516}, respectively. For the reader's convenience, the most relevant results or this prior research are summarized here:
%\begin{itemize}
 %   \item A swirling air flow with the right axial and azimuthal components through a round cavity can lead to a supercritical Hopf bifurcation of the first pair of spinning eigenmodes, causing the cavity to whistle.
 %   \item In the absence of external forcing, and zero azimuthal flow, the modes' spinning direction is subject only to spontaneous symmetry breaking. Imposing an explicit symmetry breaking by a bias flow breaks the modal degeneracy, causing a frequency split and destabilizing one mode while stabilizing the other.
 %   \item In the frequency range of interest, the non-linear cavity acoustics in the presence of a turbulent flow through the cavity can be modeled to good accuracy using low-order models which consider only the dominant mode pair.
%\end{itemize}

\section{Non-linear wave-mode coupling \label{Section 2}}
In this section, we describe the non-linear interference between the cavity modes and the incident wave in a circulator with three ports, highlighting the key ingredients required to achieve lossless non-reciprocal transmission. The concept and the theoretical foundations are not limited to acoustic circulators. As in \cite{Wang2005369}, cross-coupling between the modes is neglected under the assumption that the dominant mechanisms are the coupling between the incident and the outgoing wave (W$\shortrightarrow$W), coupling from the incident wave to the cavity modes (W$\shortrightarrow$M) and coupling from the cavity modes to the outgoing wave (M$\shortrightarrow$W). As discussed in the next section, only one mode ($a_-$) is undergoing a limit cycle. The non-linear coupling between the cavity modes $a=(a_+,a_-)^T\in\mathbb{C}^2$ and the incident wave $\vert s_\mathrm{in}\rangle\in\mathbb{C}^3$ is governed by the following system: 
\begin{eqnarray}
    \dot{a}_\pm&=&\underbrace{i\omega_{\pm} a_\pm }_{\text{oscillation}} -\underbrace{\gamma a_\pm}_{\text{dissipation}} +\underbrace{\frac{\beta_\pm a_\pm}{1+\kappa \vert a_\pm\vert^2}}_{\text{amplification}}  +\underbrace{D^\dag_\pm \vert s_{\mathrm{in}}\rangle}_{\text{W}\shortrightarrow \text{M}},\label{modal dynamics 2m}\\
   \vert s_{\mathrm{out}}\rangle&=&\underbrace{C \vert s_{\mathrm{in}}\rangle}_{\text{W}\shortrightarrow \text{W}} + \underbrace{D a}_{\text{M}\shortrightarrow \text{W}}, \label{input-output 2m}
\end{eqnarray}
where $\omega_{\pm}$ are the eigenfrequencies, $\gamma$ is the decay rate, $\beta_\pm\in\mathbb{R}^+$ are the linear gain coefficients, $\kappa \in \mathbb{R}^+$ is the non-linearity constant, $C\in \mathbb{R}^{3\times 3}$ is the background scattering matrix, $\vert s_\mathrm{out}\rangle\in\mathbb{C}^3$ is the outgoing wave, $D\in\mathbb{C}^{3\times 2}$ is the coupling matrix. A dot denotes the time derivative and $\dag$ is the Hermitian transpose. Explicit expressions of $D$ and $C$ are given in the Supplementary Notes \cite{SuppMat}. Note that Eq. \eqref{modal dynamics 2m} corresponds to two decoupled equations, one for $a_+$ and one for $a_-$. The eigenfrequencies $\omega_\pm$ are generally assumed to be quasi-degenerate ($\omega_+\approx \omega_-$) and well-isolated in the cavity spectrum, meaning that there are no nearby resonances. \qq{Analogously, the linear gain coefficients are assumed to be similar: $\beta_+\approx\beta_-$.} Due to symmetry, the decay rate $\gamma$ \qq{and the non-linearity coefficient $\kappa$ are} assumed to be the same for both modes. To compute the steady-state forced response of $a_\pm$, we assume that the modes are perfectly synchronized with the \cc{incident wave}. The functional form of the amplification term in Eq. \eqref{modal dynamics 2m} is motivated by experimental results \cite{Bourquard2020} and this type of saturable gain was also used in  \cite{Pedergnana2022CMT}. For a wave incident from the first port, $\vert s_\mathrm{in}\rangle=s e^{i\omega t} \vert 1\rangle$, where $s$ is the incident wave amplitude, $\omega=2\pi f$ is the angular forcing frequency and $\vert 1\rangle=(1,0,0)^T$ is a unit vector. We define the nondimensionalized amplitude as \qq{$\tilde{s}=s \sqrt{\kappa/\beta}$, where $\beta=(\beta_++\beta_-)/2$ is the arithmetic mean of the linear gain coefficients.} Using the definition of the scattering matrix $S$ given by $\vert s_\mathrm{out}\rangle=S\vert s_\mathrm{in}\rangle$, a general expression for $S$ can be found using the Moore-Penrose pseudoinverse \cite{Pedergnana2022CMT}:
\begin{eqnarray}
    S=\dfrac{\vert s_\mathrm{out}\rangle\langle s_\mathrm{in}\vert}{\langle s_\mathrm{in}\vert s_\mathrm{in}\rangle}. \label{general scattering matrixm}
\end{eqnarray}
In general, $S$ is a non-linear function of $\vert s_\mathrm{in}\rangle$ because the outgoing wave $\vert s_\mathrm{out}\rangle$ is a linear function of the cavity modes $a_\pm$, which, in turn, depend non-linearly on $\vert s_\mathrm{in}\rangle$. 

\qq{In the following, we provide analytical results which numerically illustrate the mechanism visualized in Fig. \ref{Figure 1}. We recall that, in order to synchronize with the limit cycle so as to profit from the energy of its self oscillations, the incident amplitude must be sufficiently large. In the limit of infinitely large amplitudes $\vert a_\pm\vert$, the modal equations \eqref{modal dynamics 2m} reduce to the classical temporal coupled mode theory which describes linear scattering at a resonant cavity \cite{Fan2003569,Suh20041511,Wang2005369}. This limit plays an important role in the non-linear transmission scenario we propose. One can imagine the limit of infinitely large amplitude as a dissipative ground state which we seek to overcome by synchronizing with the cavity limit cycle. In practice, the goal is to increase the incident wave amplitude to a large enough value such that the resulting scattering matrix is a weakly perturbed version of its infinite amplitude limit, and then to tune the amplitude and the modal parameters until the desired scattering state is reached. By expanding the modal dynamics \eqref{modal dynamics 2m} in a perturbation series in $s^{-1}$, one can compute the leading-order correction to the scattering matrix in the limit of large incident wave amplitudes. The derivation of the non-linear scattering matrix and its large amplitude approximation is provided in the Supplementary Notes \cite{SuppMat}. To model dissipation in our model, we introduce the unitarity factor $\sigma\in\mathbb{R}^+$, which is equal to $1$ for a \qq{perfectly} reversible process and takes values between 0 and 1 \qq{if the scattering is lossy}. Thus, $\sigma$ is a measure of the dissipation present in a real-world system that must be overcome to achieve lossless scattering. Next, we assume that the Zeeman splitting and incident wave frequency satisfy $\omega_\pm=\omega\pm\sqrt{3}\gamma\sigma/(\sigma+2)$, which corresponds to perfect matching (see the Supplementary Notes \cite{SuppMat}) Computing the synchronized forced response of the modal amplitudes $a_\pm = \vert a_\pm \vert e^{i\omega t+\varphi_\pm}$ for an incident wave from port  ``1'', $\vert s_\mathrm{in}\rangle=s e^{i \omega t}\vert 1 \rangle$, yields the amplitude transmission coefficients for large normalized amplitudes $\tilde{s}$ \cite{SuppMat}:}
\begin{eqnarray}
 T_{1\shortrightarrow 2}&\approx& \bigg \vert \dfrac{1-\sigma}{3} -\dfrac{(2\sigma^2+2\sigma-1)}{  \tilde{s}^2 (\sigma^2 + \sigma + 1)}\bigg \vert+O(\tilde{s}^{-4}), \label{T12lim4m} \\
   T_{1\shortrightarrow 3}&\approx&\bigg \vert \dfrac{ (1+2\sigma)  }{3}+\dfrac{(1+4\sigma+\sigma^2)}{ \tilde{s}^2  (\sigma^2 + \sigma + 1)}\bigg \vert+O(\tilde{s}^{-4}).  \label{T13lim4m}
\end{eqnarray}
\qq{For mildly dissipative systems with $0.4<\sigma < 1$, and large enough $\tilde{s}$, Eq. \eqref{T12lim4m} shows that the non-linear interference of the incident wave with the cavity limit cycle decreases transmission to the blocked port ``2'', eventually reaching an identically zero value. Under these conditions, Eq. \eqref{T13lim4m} indicates that transmission to port ``3'' is enhanced and reaches values close to or even exceeding unity, depending on level of dissipation $\sigma$. The above results show that qualitatively, this synchronization-based mechanism enables an increase in the non-reciprocal behavior of the system, while at the same time compensating transmission losses. Quantitative comparisons between theory and experiments will be provided in Sec. \ref{Section 4}.}

\section{Experimental validation\label{Section 3}}
\qq{For our scattering experiments, we built \qq{and characterized} a circulator which consists of \cc{a deep} annular cavity \cc{subject to} a swirling flow \cite{Abel2023_2} \cc{and is equipped with three equispaced ports along its circumference}. The bulk flow speed, the swirl number of the flow and the amplitude of the incident wave \cc{through one of the ports} can be set individually to tune the non-linear scattering matrix. Three waveguides are \cc{connected} to the cavity ports, \cc{thus forming an acoustic circulator}.} The experimental setup is shown in Fig. \ref{Figure 2}a. Technical information is given in Appendix \ref{Appendix_Exp}. Let us first consider Fig. \ref{Figure 2}a in the absence of acoustic forcing. A swirling flow is imposed in the blue pipe, and passes through \cc{the center of the axisymmetric} cavity. At suitable conditions, \cc{the} swirling jet in the cavity exhibits self-sustained oscillations thanks to the intrinsic coupling between one of its hydrodynamic modes and a spinning acoustic mode of the cavity. The trapped aeroacoustic mode producing this cavity whistling is shown in Fig. \ref{Figure 2}b, where one can see the acoustic pressure field that interacts with the coherent vorticity fluctuations. The latter fluctuations develop along the shear layer in the cavity aperture region, and were measured with particle image velocimetry (see Fig. \ref{Figure 2}c). A bifurcation diagram of the acoustic pressure inside the cavity, depicting the transition between the quiet and the whistling phase, is presented in Appendix \ref{Appendix_BifDiag}.

Once an aeroacoustic limit cycle has been established and the cavity is whistling, acoustic forcing with a specified amplitude is imposed from a compression driver mounted on one of the ports. The incident and outgoing signals are reconstructed using microphones mounted on the ports to obtain the transmission coefficients. For sufficiently strong and in-tune forcing, the cavity limit cycle can synchronize with the forcing, as demonstrated in Appendix \ref{Appendix_Sync}. Therefore, this experimental set-up is a realistic representation of the model assumptions made in Sec. \ref{Section 2}.
%\qq{An azimuthal bias flow \cc{produced with ventilators} can break the symmetry of a round circulator cavity, enabling perfect non-reciprocity when the flow speed is tuned correctly \cite{Fleury2014516}.} In \cc{\cite{Abel2023_1,Abel2023_2}}, it was shown that an axial flow passing through a \cc{deep annular} cavity can \cc{induce a spinning aeroacoustic} limit cycle \cc{involving its first azimuthal acoustic mode}, which manifests itself as a whistling tone at roughly the \cc{eigenfrequency} of one of \cc{this pure acoustic mode}. We combine the two aforementioned effects in this work to create a system featuring \cc{an explicit symmetry breaking} and a limit cycle. Only the most pertinent facts are state here; the interested reader is referred to the references for more details about the underlying physics of the scatterer used here. 

\begin{figure}[t!]%
\centering
\begin{psfrags}
 \psfrag{a}{{\text{\tiny{\fontfamily{phv}\selectfont
\hspace{-0.1cm} \textcolor{f1purp2}{Port 1}} }}}
 \psfrag{b}{{\text{\tiny{\fontfamily{phv}\selectfont
\hspace{-0.1cm} \textcolor{f1purp2}{Port 2}} }}}
 \psfrag{c}{{\text{\tiny{\fontfamily{phv}\selectfont
\hspace{-0.1cm} \textcolor{f1purp2}{Port 3}} }}}
   \psfrag{d}{\tiny{\fontfamily{phv}\selectfont
\hspace{-0.05cm}\textcolor{blue2}{}}}
   \psfrag{e}{\tiny{\fontfamily{phv}\selectfont
\hspace{0.1cm}\textcolor{f1red2}{Swirling air flow [{$\shortrightarrow$}]}}}
   \psfrag{f}{\tiny{\fontfamily{phv}\selectfont
\textcolor{green1}{$\theta$}}}
   \psfrag{g}{\tiny{\fontfamily{phv}\selectfont
\hspace{0.05cm}\textcolor{green1}{$x$}}}
   \psfrag{h}{\textbf{\tiny{\fontfamily{phv}\selectfont
a}}}
   \psfrag{A}{\textbf{\tiny{\fontfamily{phv}\selectfont
b\hspace{4.5cm}c}}}
    \psfrag{i}{\tiny{\fontfamily{phv}\selectfont
\hspace{-0.02cm}min.}}
    \psfrag{j}{\tiny{\fontfamily{phv}\selectfont
\hspace{-0.019cm}max.}}
    \psfrag{m}{\tiny{\fontfamily{phv}\selectfont
\begin{tabular}{ c }
\text{\tiny{Vorticity}}  \\ 
\text{\tiny{fluctuations}}   
\end{tabular}}}
    \psfrag{l}{\tiny{\fontfamily{phv}\selectfont
\hspace{-0.01cm}Amplitude}}
      \psfrag{+}{\tiny{\fontfamily{phv}\selectfont
\hspace{0.055cm}-1\hspace{0.18cm}1}}
      \psfrag{-}{\tiny{\fontfamily{phv}\selectfont
\hspace{-0.02cm}$\tilde{\Omega}_\theta$ [s\textsuperscript{-1}]}}
 \psfrag{n}{{\text{\tiny{\fontfamily{phv}\selectfont
\hspace{0.01cm} \textcolor{f2blue}{Source}} }}}
 \psfrag{B}{{\text{\tiny{\fontfamily{phv}\selectfont
\hspace{-0cm} {70 cm}} }}}
\psfrag{Z}{{\text{\tiny{\fontfamily{phv}\selectfont
\hspace{-0.5cm} cylindrical cavity} }}}
  \psfrag{?}{\tiny{\fontfamily{phv}\selectfont
flow direction}}
\includegraphics[width=0.68\textwidth]{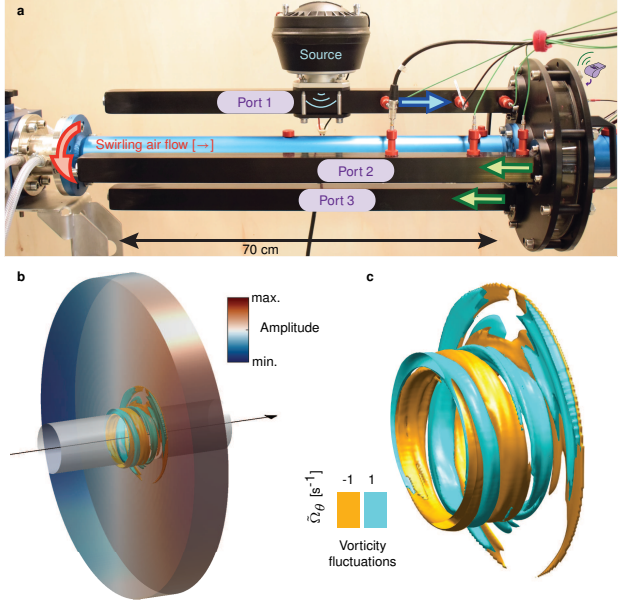}
\end{psfrags}
\caption{\textbf{Acoustic circulator based on a spinning cavity limit cycle}. \textbf{a} Experimental setup. A swirling flow with fixed axial and azimuthal components is imposed in the blue pipe which passes through a cylindrical cavity, leading to a limit cycle (whistling) \cc{at around $800$ Hz}. In contrast with classic wind instruments whose sound radiation originates from self-oscillating standing modes, our special whistle involves spinning modes. For the scattering experiments, three 70 cm long square waveguides are \cc{connected} to the \cc{upstream wall of} the round cavity \cc{at equispaced angular positions ($120^\circ$ apart)}. \cc{These waveguides are equipped with sound absorbing foam at their left termination in order to provide quasi-anechoic conditions.} For the scattering experiments, incident waves are sent to the cavity using a compression driver (source) \cc{mounted on one of the waveguides}. Microphones in the waveguides are used to reconstruct the incident and outgoing waves. \textbf{b} Acoustic mode of the circulator cavity, also showing isosurfaces of the azimuthal component of the coherent vorticity fluctuations \cc{$\tilde{\Omega}_\theta$} (see Appendix \ref{Appendix_AcField} and \ref{Appendix_Fluid}). \textbf{c} Zoomed-in view of this instantaneous snapshot of the vorticity fluctuations, which was obtained from stereoscopic particle image velocimetry.}\label{Figure 2}
\end{figure}

\section{Results\label{Section 4}}
\begin{figure}[t]%
\centering
\begin{psfrags}
    \psfrag{l}{\tiny{\fontfamily{phv}\selectfont
\hspace{-0.09cm}8}}
    \psfrag{9}{\tiny{\fontfamily{phv}\selectfont
\hspace{-0.05cm}6}}
    \psfrag{8}{\tiny{\fontfamily{phv}\selectfont
\hspace{-0.05cm}4}}
    \psfrag{7}{\tiny{\fontfamily{phv}\selectfont
\hspace{-0.05cm}2}}
    \psfrag{o}{\tiny{\fontfamily{phv}\selectfont
\hspace{-0.5cm}transmission}}
    \psfrag{k}{\tiny{\fontfamily{phv}\selectfont
\hspace{-0.05cm}0}}
    \psfrag{i}{\tiny{\fontfamily{phv}\selectfont
\hspace{-0.1cm}780}}
  \psfrag{3}{\tiny{\fontfamily{phv}\selectfont
\hspace{-0.1cm}810}}
    \psfrag{j}{\tiny{\fontfamily{phv}\selectfont
\hspace{-0.2cm}820}}
    \psfrag{n}{\tiny{\fontfamily{phv}\selectfont
\hspace{-1.1cm}frequency (Hz)}}
    \psfrag{m}{\tiny{\fontfamily{phv}\selectfont
\hspace{0.1cm}$ s_\mathrm{in} $ (Pa)}}
    \psfrag{e}{\tiny{\fontfamily{phv}\selectfont
\hspace{-0cm}10}}
    \psfrag{4}{\tiny{\fontfamily{phv}\selectfont
\hspace{-0cm}37}}
    \psfrag{5}{\tiny{\fontfamily{phv}\selectfont
\hspace{-0cm}65}}
    \psfrag{6}{\tiny{\fontfamily{phv}\selectfont
\hspace{-0cm}93}}
    \psfrag{b}{\tiny{\fontfamily{phv}\selectfont
\hspace{-0.1cm}120}}
    \psfrag{f}{\tiny{\fontfamily{phv}\selectfont
\hspace{-0.1cm}8}}
    \psfrag{d}{\tiny{\fontfamily{phv}\selectfont
\hspace{-0.08cm}0}}
    \psfrag{c}{\tiny{\fontfamily{phv}\selectfont
\hspace{-0.43cm}transmission}}
   \psfrag{g}{\tiny{\fontfamily{phv}\selectfont
10\hspace{0.4cm} 120}}
   \psfrag{a}{\tiny{\fontfamily{phv}\selectfont
\hspace{-0.2cm}$s_\mathrm{in}$ (Pa)}}
   \psfrag{h}{\tiny{\fontfamily{phv}\selectfont
max}}
   \psfrag{J}{\tiny{\fontfamily{phv}\selectfont
1$\shortrightarrow$3}}
   \psfrag{Q}{\tiny{\fontfamily{phv}\selectfont
1$\shortrightarrow$2}}
   \psfrag{H}{\tiny{\fontfamily{phv}\selectfont
100$\%$}}
   \psfrag{s}{\tiny{\fontfamily{phv}\selectfont
\hspace{-0.25cm}min}}
   \psfrag{r}{\tiny{\fontfamily{phv}\selectfont
\hspace{-0.3cm}max}}
   \psfrag{y}{\tiny{\fontfamily{phv}\selectfont
\hspace{-1.0cm}incident wave amp. $ {s}_\mathrm{in} $ (Pa)}}
   \psfrag{x}{\tiny{\fontfamily{phv}\selectfont
\hspace{-0.07cm}4}}
   \psfrag{T}{\tiny{\fontfamily{phv}\selectfont
\hspace{-0.07cm}2}}
   \psfrag{s}{\tiny{\fontfamily{phv}\selectfont
\hspace{-0.07cm}0}}
   \psfrag{0}{\tiny{\fontfamily{phv}\selectfont
\hspace{-0.07cm}8}}
   \psfrag{p}{\tiny{\fontfamily{phv}\selectfont
\hspace{-0.07cm}6}}
   \psfrag{q}{\tiny{\fontfamily{phv}\selectfont
820 }}
   \psfrag{v}{\tiny{\fontfamily{phv}\selectfont
\hspace{-0.8cm}transmission 1$\shortrightarrow$3 (dB)}}
   \psfrag{u}{\tiny{\fontfamily{phv}\selectfont
18 }}
   \psfrag{S}{\tiny{\fontfamily{phv}\selectfont
8.1 }}
   \psfrag{R}
   {\tiny{\fontfamily{phv}\selectfont
-1.4}}
   \psfrag{q}{
   \tiny{\fontfamily{phv}\selectfont
\hspace{-0.2cm}2 }}
   \psfrag{P}{\tiny{\fontfamily{phv}\selectfont
\hspace{-0.2cm}1.5 }}
  \psfrag{B}{\tiny{\fontfamily{phv}\selectfont
\hspace{-0.1cm}1 }}
   \psfrag{h}{\tiny{\fontfamily{phv}\selectfont
\hspace{-0.2cm}0.5}}
   \psfrag{I}{\tiny{\fontfamily{phv}\selectfont
\hspace{-0.1cm}0}}
   \psfrag{M}{\tiny{\fontfamily{phv}\selectfont
-11 }}
   \psfrag{t}{\tiny{\fontfamily{phv}\selectfont
-20 \hspace{0.25cm} -11 \hspace{0.25cm} -1.4  \hspace{0.25cm} 8.1 \hspace{0.25cm} 18 }}
   \psfrag{F}{\tiny{\fontfamily{phv}\selectfont
10 }}
   \psfrag{W}{\tiny{\fontfamily{phv}\selectfont
-3.2 }}
   \psfrag{V}{\tiny{\fontfamily{phv}\selectfont
-16 }}
   \psfrag{U}{\tiny{\fontfamily{phv}\selectfont
-29 }}
\psfrag{E}{\tiny{\fontfamily{phv}\selectfont
-41 \hspace{0.25cm} -29 \hspace{0.25cm} -16  \hspace{0.25cm} -3.2 \hspace{0.25cm} 10 }}
   \psfrag{G}{\tiny{\fontfamily{phv}\selectfont
\hspace{-0.5cm}isolation (dB) }}
   \psfrag{Z}{\tiny{\fontfamily{phv}\selectfont
$s_\mathrm{in}$=100 Pa }}
   \psfrag{Y}{\tiny{\fontfamily{phv}\selectfont
$s_\mathrm{in}$=50 Pa  }}
   \psfrag{X}{\tiny{\fontfamily{phv}\selectfont
$s_\mathrm{in}$=10 Pa}}
   \psfrag{w}{\tiny{\fontfamily{phv}\selectfont
\hspace{-0.02cm}2 }}
  \psfrag{z}{\tiny{\fontfamily{phv}\selectfont
\hspace{-0.03cm}0 }}
  \psfrag{D}{\tiny{\fontfamily{phv}\selectfont
\hspace{-0.5cm}transmission }}
  \psfrag{A}{\tiny{\fontfamily{phv}\selectfont
780 \hspace{0.19cm} 790\hspace{0.19cm} 800
\hspace{0.19cm} 810\hspace{0.19cm} 820}}
  \psfrag{C}{\tiny{\fontfamily{phv}\selectfont
\hspace{-0.45cm}frequency (Hz) }}
  \psfrag{K}{\tiny{\fontfamily{phv}\selectfont
0 }}
  \psfrag{L}{\tiny{\fontfamily{phv}\selectfont
8 }}
\psfrag{N}{\textbf{\text{\tiny{\fontfamily{phv}\selectfont
\hspace{-0.1cm}a }}}\hspace{3.5cm}
\textbf{\text{\tiny{\fontfamily{phv}\selectfont
b }}}\hspace{3.5cm}
\textbf{\text{\tiny{\fontfamily{phv}\selectfont
c }}}}
\psfrag{O}{\textbf{\text{\tiny{\fontfamily{phv}\selectfont
\hspace{-0.1cm}d }}}\hspace{3.5cm}
\textbf{\text{\tiny{\fontfamily{phv}\selectfont
e }}}\hspace{3.5cm}
\textbf{\text{\tiny{\fontfamily{phv}\selectfont
f }}}}
   \psfrag{1}{\tiny{\fontfamily{phv}\selectfont
\textbf{Experiments} \hspace{2.2cm} \textbf{Theory} }}
   \psfrag{2}{\tiny{\fontfamily{phv}\selectfont
\hspace{0.6cm}increasing excitation amplitude}}
   \psfrag{X}{\tiny{\fontfamily{phv}\selectfont
\hspace{-0.35cm}$\Delta s =0$}}
   \psfrag{Y}{\tiny{\fontfamily{phv}\selectfont
\hspace{-0.45cm}$\Delta s =0.36$}}
   \psfrag{Z}{\tiny{\fontfamily{phv}\selectfont
\hspace{-0.45cm}$\Delta s =0.82$}}
   \psfrag{3}{\tiny{\fontfamily{phv}\selectfont
\hspace{-0.35cm}$\Delta \tilde{s} =0$}}
   \psfrag{4}{\tiny{\fontfamily{phv}\selectfont
\hspace{-0.45cm}$\Delta \tilde{s} =0.36$}}
   \psfrag{5}{\tiny{\fontfamily{phv}\selectfont
\hspace{-0.45cm}$\Delta \tilde{s} =0.82$}}
      \psfrag{7}{\tiny{\fontfamily{phv}\selectfont \textcolor{f3blue}{$T_{1\shortrightarrow 2}$}}}
      \psfrag{8}{\tiny{\fontfamily{phv}\selectfont \textcolor{black}{$T_{1\shortrightarrow 3}$} }}
\includegraphics[width=0.78\textwidth]{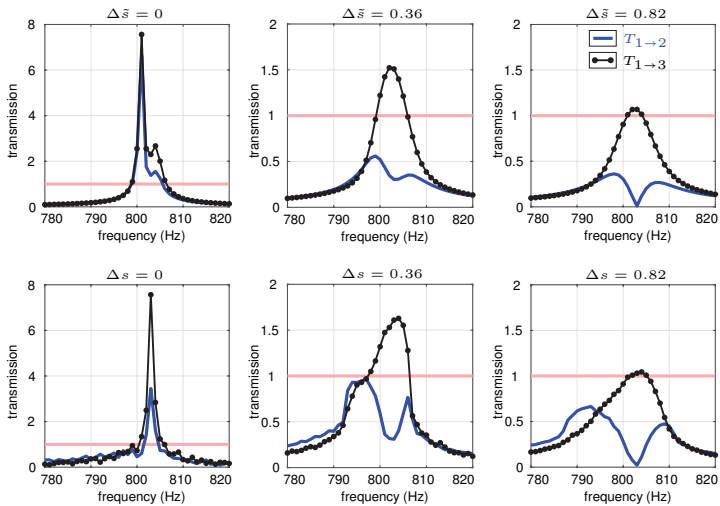}
\end{psfrags}
\caption{\textbf{Experimental validation of a perfect lossless circulator}. Amplitude transmission coefficients for increasing incident wave amplitudes $\Delta \tilde{s}$ and $\Delta s$ (expressed relative the intervals $\tilde{s}\in[0.04,2.1]$ and $s\in [10,120]$ Pa, respectively; see Sec. \ref{Section 4}). Top row: Theoretical results obtained from the non-linear coupled-mode theory described in Sec. \ref{Section 2}. Bottom row: experimental measurements performed on the setup described in Sec. \ref{Section 3}. In both cases, a harmonic incident wave is imposed in port ``1''. The blue and black curves refer to the transmitted wave amplitudes at ports ``2'' and ``3'', respectively, relative to the incident wave amplitude and frequency.}\label{Figure 3}
\end{figure}

In order to facilitate the comparison of theoretical and experimental results, we define the incident wave amplitude relative to the maximum range of amplitudes $[s_\mathrm{min},s_\mathrm{max}]$, imposed in the scattering process: $\Delta s=(s-s_\mathrm{min})/(s_\mathrm{max}-s_\mathrm{min})$, and analogously for $\Delta \tilde {s}$. We show in Fig. \ref{Figure 3} that our theoretical model reproduces the non-linear scattering experiments with remarkable accuracy using only few parameters. The ranges of excitation amplitudes used in the model and in the experiments were $\tilde{s}\in[0.04,2.1]$ and  $s\in [10,120]$ Pa, respectively. No fit was performed to adjust the model parameters, but they were manually defined to match the experiments. The parameter values used in the model were $\sigma=0.61$, $\gamma=27.5$ s\textsuperscript{-1}, $f_+=\omega_+/2\pi=804.3$ Hz, $f_-=\omega_-/2\pi=800.8$ Hz, $\beta_+=21.2$ s\textsuperscript{-1}, $\beta_-=30.3$ s\textsuperscript{-1}, $\kappa=1$ (in units of $[a^{-2}]$). For these values, perfect matching in the model is achieved at 802.5 Hz, similar to what is observed in the experiments (see also Appendix \ref{Appendix_Zeeman}).

For small incident wave amplitudes, sharp peaks are visible because the self-sustained radiation of the limit cycle dominates transmission through the ports. As the incident wave amplitude is increased, a non-reciprocal, lossless scattering state emerges. To verify the robustness of the experimental results shown in Fig. \ref{Figure 3}, we investigated the sensitivity of the lossless, non-reciprocal state (see Appendix \ref{Appendix_Unbiased}). For this, we repeated the scattering experiments for a fixed value of $\Delta s=9$ while varying the axial and azimuthal components of the swirling flow. The results show that there exists a continuum of low-loss, non-reciprocal scattering states in the vicinity of the operating point treated in this study. Therefore, our measurements evidence the relevance of loss-compensated non-reciprocity as a robust synchronisation phenomenon which persists under small variations of the operating conditions. As limit cycles are ubiquitous in dynamical systems with saturable gain, we surmise that synchronization-based lossless non-reciprocity may be transposed to other physical platforms, from electronics to non-Hermitian photonics. Semiconductor-based electronic amplifiers and continuous optical gain media may allow for a direct implementation of our concept in high-frequency circuits or silicon nanophotonics. More fundamentally, our loss-compensation mechanism may open new experimental opportunities in topological physics, where time-reversal symmetry plays a pivotal role in the robustness of chiral edge states, which can be induced in circulator networks \cite{Zhang2021293,Zhang2023SCIADV}. The complex interplay between non-linearity and gain to produce limit cycles capable of breaking reciprocity remains a promising avenue, and we envision that extensions to periodic lattices, or systems coupled to a radiation continuum may lead to a wealth of unforeseen progress in a field that used to be thought as fundamentally hindered by absorption losses.


\subsection*{Data availability}
The data that support the findings of this study are available at \url{https://polybox.ethz.ch/index.php/s/CLxLxoDsWkSuIgq}.
\subsection*{Code availability}
The codes that support the findings of this study are available at \url{https://polybox.ethz.ch/index.php/s/CLxLxoDsWkSuIgq}.

\subsection*{Contributions}
\noindent N.N. conceptualized and supervised the research. T.P. and A.F. conducted the acoustic and flow measurements. T.P. and N.N. developed the theory. T.P. performed data analysis and validated the model. T.P., A.F., R.F., and N.N. discussed the results. T.P. wrote the first draft of the article. N.N. and R.F. reviewed and edited it. All authors approved the final version of the article.

\bibliography{sn-bibliography}% common bib file
%\input sn-article.bbl

%\label{Appendix_Exp}
%\label{Appendix_AcField}
%\label{Appendix_Fluid}
%\label{Appendix_BifDiag}
%\label{Appendix_Zeeman}
%\label{Appendix_Unbiased}
%\label{Appendix_Sync}

\renewcommand\thesubsection{\Alph{subsection}}
\section*{Appendix}
\setcounter{subsection}{0}
\subsection{Scattering experiments}
\label{Appendix_Exp}
\begin{figure}[h]%
\centering
\begin{psfrags}
    \psfrag{a}{\tiny{\fontfamily{phv}\selectfont
\textcolor{red}{microphones}}}
    \psfrag{D}{\tiny{\fontfamily{phv}\selectfont
\textcolor{red}{microphones}}}
    \psfrag{b}{\tiny{\fontfamily{phv}\selectfont
\textcolor{green1}{$x$}}}
    \psfrag{c}{\tiny{\fontfamily{phv}\selectfont
$R$=12.8 cm}}
    \psfrag{d}{\tiny{\fontfamily{phv}\selectfont
\hspace{-0.1cm}$D$=4 cm}}
    \psfrag{e}{\hspace{-0.05cm}\tiny{\fontfamily{phv}\selectfont
4.6 cm}}
    \psfrag{f}{\tiny{\fontfamily{phv}\selectfont
120$ ^{\circ}$}}
    \psfrag{B}{\tiny{\fontfamily{phv}\selectfont
120$ ^{\circ}$}}
    \psfrag{g}{\tiny{\fontfamily{phv}\selectfont
\textcolor{green1}{$x$}}}
    \psfrag{h}{\tiny{\fontfamily{phv}\selectfont
\hspace{-0.15cm}{air blower}}}%\textcolor{SIorange}
    \psfrag{i}{\tiny{\fontfamily{phv}\selectfont
{shop air}}}%\textcolor{SIorange}
    \psfrag{i}{\tiny{\fontfamily{phv}\selectfont
{shop air}}}%\textcolor{SIorange}
    \psfrag{j}{\tiny{\fontfamily{phv}\selectfont
\textcolor{SIblue}{mass flow meter}}}
   \psfrag{C}{\tiny{\fontfamily{phv}\selectfont
\textcolor{SIblue}{mass flow meter}}}
    \psfrag{k}{\tiny{\fontfamily{phv}\selectfont
outlet}}
    \psfrag{l}{\tiny{\fontfamily{phv}\selectfont
\textcolor{SIbrown}{compression driver (source)}}}
    \psfrag{m}{\tiny{\fontfamily{phv}\selectfont
acoustic foam}}
    \psfrag{n}{\tiny{\fontfamily{phv}\selectfont
70 cm}}
    \psfrag{o}{\tiny{\fontfamily{phv}\selectfont
\textcolor{SIpurple}{swirler}}}
    \psfrag{A}{\tiny{\fontfamily{phv}\selectfont
3 cm}}
    \psfrag{E}{\tiny{\fontfamily{phv}\selectfont
3.2 cm}}
    \psfrag{F}{\tiny{\fontfamily{phv}\selectfont
1}}
    \psfrag{G}{\tiny{\fontfamily{phv}\selectfont
2}}
    \psfrag{H}{\tiny{\fontfamily{phv}\selectfont
3}}
   \psfrag{I}{\tiny{\fontfamily{phv}\selectfont
\textcolor{green1}{$\theta$}}}
   \psfrag{Z}{\tiny{\fontfamily{phv}\selectfont
\textcolor{green1}{$r$}}}
   \psfrag{K}{\large{\fontfamily{phv}\selectfont
thermocouple}}
   \psfrag{>}{}%{\large{\fontfamily{phv}\selectfont
%\textcolor{red}{$\dots$}}}
   \psfrag{L}{\tiny{\fontfamily{phv}\selectfont
\hspace{-0.05cm}\textcolor{armorange}{semi-anechoic waveguides}}}
   \psfrag{J}{\tiny{\fontfamily{phv}\selectfont
{cylindrical cavity}}}
   \psfrag{M}{\tiny{\fontfamily{phv}\selectfont
{anechoic end \hspace{4.8cm} anechoic end}}}
   \psfrag{K}{\tiny{\fontfamily{phv}\selectfont
\textcolor{swirlpurp}{swirling air flow}}}
   \psfrag{N}{\tiny{\fontfamily{phv}\selectfont
front view \hspace{2.05cm} back view}}
   \psfrag{<}{\tiny{\fontfamily{phv}\selectfont
\hspace{-0.1cm}side view}}
   \psfrag{1}{\tiny{\fontfamily{phv}\selectfont
\textcolor{Vblue}{$V$}}}
   \psfrag{2}{\tiny{\fontfamily{phv}\selectfont
\hspace{0.2cm}\textcolor{Ured}{$U$}}}
\includegraphics[width=0.7\textwidth]{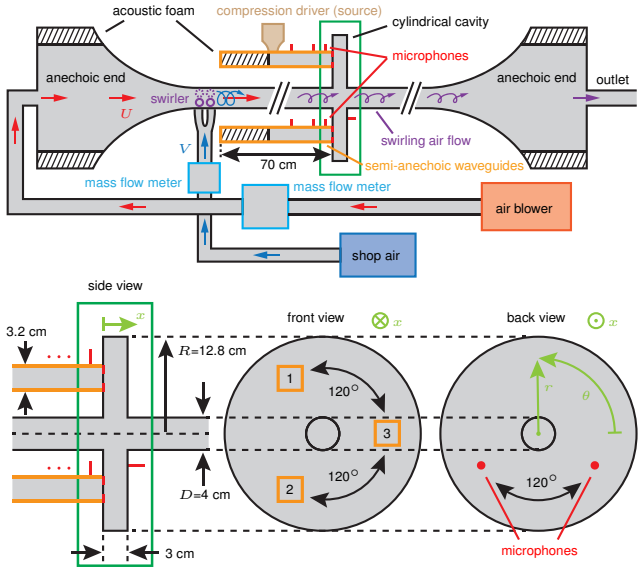}
\end{psfrags}
\caption{Sketch of the experimental setup used in this work (not for scale). A blower and shop air from the laboratory supply \cc{the wind channel of the experimental setup. The respective air flows} are monitored using mass flow meters. \cc{At the swirler, air is injected tangentially to the wall of the circular cross-section wind channel, and it mixes with the main air from the blower. This configuration enables full control of the swirl intensity of the jet flowing through the cavity. This deep axisymmetric cavity is the central element of the 3-port circulator, which is equipped with} three waveguides. \cc{Their angular positions are $120^\circ$ apart at the same radius. One of the waveguides} is equipped with a compression driver to acoustically \cc{force one of the three circulator's ports.} Microphones are placed \cc{along} the \cc{waveguides} and on the cavity for acoustic measurements. The wind channel has anechoic, catenoidal terminations which are filled with acoustic foam. Similarly, the ends of the waveguides attached to the cavity are also filled with anoechoic foam to achieve low \cc{acoustic} reflection.}\label{Appendix Figure 2}
\end{figure}

To test our theoretical predictions, we designed an aeroacoustic circulator, shown in Fig. \ref{Figure 2}a, which uses a low-Mach air flow to generate acoustic limit cycles. For this, a swirling flow with bulk speed $\overline{U}=U+V$, whose axial ($U$) and azimuthal ($V$) components can be separately tuned (see Figure \ref{Appendix Figure 2}), is imposed in \cc{the} pipe \cc{to which the cavity is connected}. The axial velocity $U$ is deduced from the mass ﬂow and the temperature in the channel, which are measured with a mass flow meter and a thermocouple. The azimuthal flow is induced using a swirler \cc{based on tangential injection of air in the upststream part of the pipe}, and also monitored using a mass flow meter. Note that the velocity $V$ corresponds to the contribution of the mass flow through the swirler $\dot{m}_\mathrm{s}$ to the bulk flow in the pipe, and not to the local velocity in the swirler: $V=4\dot{m}_\mathrm{s}/\rho_0 \pi D^2$, where $\rho=1.15$ kg/m\textsuperscript{3} is the ambient density and $D=4$ cm is the pipe diameter. The ambient speed of sound in the experiments is equal to $c=343$ m/s. Three \cc{semi-anechoic} waveguides (arms) with square cross-sections are connected to the cavity through compact circular holes. The ends of the arms are filled with acoustic foam and have been measured to reflect less than $13\%$ of \cc{the} acoustic energy for all considered operating conditions. Three G.R.A.S. 46BD $1/4$'' CCP microphones are flush mounted in each of the arms $x=-4.6$ cm, $x=-12.1$ cm and $x=-24.6$ cm, where $x$ is measured from the upstream inner cavity wall. The multi-microphone method was used to reconstruct the traveling waves in the arms at the incident wave frequency from the acoustic measurements \cite{Jang19981520}. Two microphones of the same type as in the \cc{waveguides} were flush mounted on the downstream wall of the cavity at the same radial positions $r=9$ cm, separated by 120$^\circ$ degrees, to measure the acoustic limit cycle oscillations in the absence of forcing \cc{from any of the waveguides}. The incident waves were imposed by a Beyma CP-850Nd compression driver mounted on one of the arms. The voltage of the signal to the drivers can be calibrated at each frequency to achieve a specified acoustic amplitude $s_\mathrm{in}$.

\subsection{Acoustic field}
\label{Appendix_AcField}
The acoustic mode shown in Fig. \ref{Figure 2}b corresponds to the analytical solution of the Helmholtz equation for a circular cavity, $J_1(r Z_1/R)\cos\theta$, where $J_1$ is the Bessel function of the first kind of order $1$, $r$ is the radial coordinate and $Z_1$ is the first zero of the first derivative of $J_1$ with respect to $\zeta$ \cite{Abel2023_2}. 

\subsection{Fluid-dynamical data} 
\label{Appendix_Fluid}
Shown in Fig. \ref{Figure 2}c are isosurfaces of the azimuthal component of the vorticity fluctuations \cc{$\tilde{\Omega}_\theta$} at the acoustic frequency, which were extracted using \cc{stereoscopic} particle image velocimetry (PIV) in an earlier study of a limit cycle at $U=25.9$ m/s and $V=0$, and which are included here for illustration purposes. The isosurfaces were produced from postprocessed data originally presented in \cite[Fig. 6a]{Abel2023_1}. The \cc{type of} aeroacoustic instability \cc{used here is based on a low Mach flow} \cite{Howe1980407,Elder1982532,Bourquard2020}, \cc{and as in the case of human whistling, the subtle interplay between the acoustic field and the rotational fluid motion results} in finite-amplitude acoustic pressure oscillations in the cavity. \cc{To engineer our circulator, we designed a special whistle in the sense that its eigenmodes are spinning inside the cavity \cite{Abel2023_1}, in contrast to the classic police whistle or wind instruments like flutes, in which standing acoustic oscillations are found. Furthermore, by imposing a swirl on the cavity flow, we can provide self-sustained aeroacoustic waves that spin against the swirl, be it in the clockwise or in the counterclockwise direction, as explained in \cite{Abel2023_2}.}

\subsection{Bifurcation diagram}
\label{Appendix_BifDiag}
\begin{figure}[h]%
\centering
\begin{psfrags}
    \psfrag{a}{\tiny{\fontfamily{phv}\selectfont
\hspace{0.145cm}2 \hspace{0.39cm} 3}}
  \psfrag{c}{\tiny{\fontfamily{phv}\selectfont
\hspace{0.06cm}1}}
  \psfrag{d}{\tiny{\fontfamily{phv}\selectfont
\hspace{-0.05cm}1$\shortrightarrow$3}}
  \psfrag{B}{\tiny{\fontfamily{phv}\selectfont
\textcolor{blue2}{$\omega$}}}
  \psfrag{v}{\tiny{\fontfamily{phv}\selectfont
\hspace{-0.35cm}\textcolor{blue2}{\hspace{0.1cm}$s_\mathrm{in} $}}}
\psfrag{y}{}%{\hspace{-7.7cm}\textbf{\text{\tiny{\fontfamily{phv}\selectfont
%a }}}\hspace{3.6cm}
%\textbf{\text{\tiny{\fontfamily{phv}\selectfont
%b }}}\hspace{3.6cm}
%\textbf{\text{\tiny{\fontfamily{phv}\selectfont
%c }}}}
\psfrag{z}{\textbf{\text{\tiny{\fontfamily{phv}\selectfont
b }}}\hspace{3.1cm}
\textbf{\text{\tiny{\fontfamily{phv}\selectfont
e }}}\hspace{3.1cm}
\textbf{\text{\tiny{\fontfamily{phv}\selectfont
h }}}}
\psfrag{A}{\textbf{\text{\tiny{\fontfamily{phv}\selectfont
c }}}\hspace{3.1cm}
\textbf{\text{\tiny{\fontfamily{phv}\selectfont
f }}}\hspace{3.1cm}
\textbf{\text{\tiny{\fontfamily{phv}\selectfont
j }}}}
  \psfrag{e}{\tiny{\fontfamily{phv}\selectfont
\hspace{-0.05cm}1$\shortrightarrow$2}}
  \psfrag{g}{\footnotesize{\fontfamily{phv}\selectfont
\hspace{0.02cm}\textcolor{purple1}{loss}}}
  \psfrag{h}{\footnotesize{\fontfamily{phv}\selectfont
\textcolor{orangex}{gain}}}
  \psfrag{f}{\footnotesize{\fontfamily{phv}\selectfont
\hspace{0cm}\textcolor{blue2}{damping}}}
  \psfrag{m}{\tiny{\fontfamily{phv}\selectfont
\hspace{-0.16cm}\textcolor{blue2}{$\vert a_0\vert$}}}
  \psfrag{r}{\tiny{\fontfamily{phv}\selectfont
\hspace{-0.78 cm}\textcolor{blue2}{$-\vert a_{\tiny 0}\vert$}\hspace{0.15cm}\textcolor{blue2}{$0$}\hspace{0.15cm}\textcolor{blue2}{$\vert a_0 \vert$ }}}
  \psfrag{j}{\tiny{\fontfamily{phv}\selectfont
\textcolor{blue2}{1} }}
  \psfrag{l}{\tiny{\fontfamily{phv}\selectfont
\textcolor{blue2}{1} }}
  \psfrag{k}{\tiny{\fontfamily{phv}\selectfont
\hspace{-0.03cm}\textcolor{blue2}{$\gamma$} }}
  \psfrag{n}{\tiny{\fontfamily{phv}\selectfont
\hspace{-0.04cm}\textcolor{blue2}{$\vert a_\pm \vert$ }}}
  \psfrag{i}{\tiny{\fontfamily{phv}\selectfont
\textcolor{blue2}{$\beta_-$} }}
  \psfrag{s}{\tiny{\fontfamily{phv}\selectfont
\hspace{-0.04cm}\textcolor{blue2}{$\vert a_\pm \vert$} }}
  \psfrag{q}{\tiny{\fontfamily{phv}\selectfont
\hspace{-0.6cm} \textcolor{blue2}{$\gamma-\beta_-$}} }
  \psfrag{p}{\tiny{\fontfamily{phv}\selectfont
\hspace{-0.6cm} \textcolor{blueweak}{$\gamma-\beta_+$}} }
  \psfrag{+}{\tiny{\fontfamily{phv}\selectfont
\hspace{0cm}\textcolor{blue2}{$0$}} }
  \psfrag{o}{\tiny{\fontfamily{phv}\selectfont
\hspace{-0.02cm}\textcolor{blue2}{$\gamma$} } }
  \psfrag{C}{\large{\fontfamily{phv}\selectfont
\hspace{0.05cm}$\dots$} }
  \psfrag{W}{\large{\fontfamily{phv}\selectfont
\hspace{0.05cm}$\dots$} }
  \psfrag{Z}{\large{\fontfamily{phv}\selectfont
\hspace{0.05cm}$\dots$} }
  \psfrag{H}{\tiny{\fontfamily{phv}\selectfont
\hspace{-0.05cm}100$\%$} }
  \psfrag{U}{\tiny{\fontfamily{phv}\selectfont
\hspace{-0.2cm} \textcolor{purple2}{$a_\pm$}} }
  \psfrag{<}{\tiny{\fontfamily{phv}\selectfont
\hspace{-0.23cm}\textcolor{purple2}{$\vert s_\mathrm{in}\rangle$} }}
  \psfrag{>}{\tiny{\fontfamily{phv}\selectfont
\hspace{-0.07cm}\textcolor{purple2}{$\vert s_\mathrm{out}\rangle$}}}
  %\psfrag{t}{\tiny{\fontfamily{phv}\selectfont
%\textcolor{blue2}{$R_\mathrm{eff}$} }}
\psfrag{t}{\footnotesize{\fontfamily{phv}\selectfont
\textcolor{blue2}{resistance} }}
  \psfrag{?}{\tiny{\fontfamily{phv}\selectfont
\hspace{-0.1cm}\textcolor{orangex}{$-$}}}
  \psfrag{-}{\tiny{\fontfamily{phv}\selectfont
\hspace{-0.05cm}\textcolor{orangex1}{$+$}}}
  \psfrag{:}{\tiny{\fontfamily{phv}\selectfont
\hspace{-0.1cm}\textcolor{redweak}{$+$}}}
  \psfrag{!}{\tiny{\fontfamily{phv}\selectfont
\hspace{-0.1cm}\textcolor{red2}{$-$}}}
  \psfrag{;}{\tiny{\fontfamily{phv}\selectfont
\hspace{-0.42cm}\textcolor{blue2}{$T_{j\shortrightarrow i}$}}}
  \psfrag{*}{\tiny{\fontfamily{phv}\selectfont
\hspace{0.1cm}\textcolor{blue2}{$\omega_\pm$}}}
\includegraphics[width=0.6\textwidth]{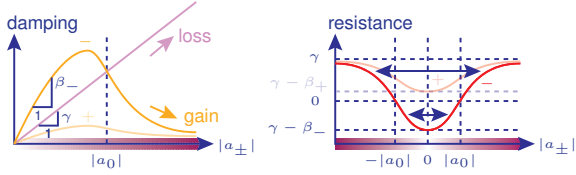}
\end{psfrags}
\caption{\cc{Some features of the theoretical model presented in the main text. Left inset:} Typical damping non-linearity of the cavity modes considered in this work. In the absence of an incident wave, if the linear gain of the ``$-$'' mode, $\beta_-$, exceeds the \cc{linear damping} $\gamma$, \cc{and if the non-linear component of the cavity gain decreases for increasing amplitude $\vert a_-\vert$,} a stable limit cycle with amplitude $\vert a_0 \vert$ occurs, \cc{which corresponds to an equilibrium between gain and loss mechanisms. In this example, the mode $a_+$ is linearly stable because $\beta_+<\gamma$.} \cc{Right inset:} Sketch of the effective resistance \cc{associated with} the non-linear cavity modes considered in this work. At large amplitudes $\vert a_\pm \vert$, the system \cc{exhibits a finite resistive term tending to $\gamma$}, i.e., its response is essentially equivalent to that \cc{of} a linearly stable resonance. }\label{Appendix Figure 1}
\end{figure}

\begin{figure}[h]%
\centering
\begin{psfrags}
    \psfrag{a}{\tiny{\fontfamily{phv}\selectfont
\hspace{-0.02cm}min}}
    \psfrag{c}{\tiny{\fontfamily{phv}\selectfont
\hspace{-0.02cm}max}}
    \psfrag{b}{\tiny{\fontfamily{phv}\selectfont
\hspace{-0.02cm}0}}
    \psfrag{d}{\tiny{\fontfamily{phv}\selectfont
\hspace{-0.5cm}ac. pressure}}
    \psfrag{e}{\tiny{\fontfamily{phv}\selectfont
0 \hspace{0.7cm} 0.5 \hspace{0.7cm} 1
 \hspace{0.7cm} 1.5 \hspace{0.7cm} 2}}
    \psfrag{n}{\tiny{\fontfamily{phv}\selectfont
\hspace{-0.25cm}27.1}}
    \psfrag{m}{\tiny{\fontfamily{phv}\selectfont
\hspace{-0.25cm}25.4}}
    \psfrag{l}{\tiny{\fontfamily{phv}\selectfont
\hspace{-0.35cm}23.6}}
    \psfrag{k}{\tiny{\fontfamily{phv}\selectfont
\hspace{-0.3cm}21.9}}
    \psfrag{j}{\tiny{\fontfamily{phv}\selectfont
\hspace{-0.3cm}20.1}}
    \psfrag{p}{\footnotesize{\fontfamily{phv}\selectfont
\hspace{-1.1cm}bulk flow speed (m/s)}}
    \psfrag{I}{\footnotesize{\fontfamily{phv}\selectfont
\hspace{-0.8cm}swirl number (-)}}
    \psfrag{4}{\tiny{\fontfamily{phv}\selectfont
\hspace{-0.05cm}$\Sigma$}}
    \psfrag{3}{\footnotesize{\fontfamily{phv}\selectfont
\hspace{-1.35cm}whistling intensity  (hPa\textsuperscript{2})}}
    \psfrag{t}{\tiny{\fontfamily{phv}\selectfont
0\hspace{0.2cm}0.9}}
    \psfrag{t}{\tiny{\fontfamily{phv}\selectfont
0\hspace{0.28cm}0.9}}
    \psfrag{w}{\tiny{\fontfamily{phv}\selectfont
$\overline{U}$}}
    \psfrag{j}{\tiny{\fontfamily{phv}\selectfont
\hspace{-0.28cm}20.1}}
    \psfrag{u}{\tiny{\fontfamily{phv}\selectfont
\hspace{-0.28cm}24.3}}
    \psfrag{f}{\tiny{\fontfamily{phv}\selectfont
\hspace{-0.1cm}0}}
    \psfrag{x}{\tiny{\fontfamily{phv}\selectfont
\hspace{-0.2cm}0.9}}
    \psfrag{5}{\tiny{\fontfamily{phv}\selectfont
0\hspace{0.57cm}0.66}}
    \psfrag{C}{\tiny{\fontfamily{phv}\selectfont
0.9}}
    \psfrag{B}{\tiny{\fontfamily{phv}\selectfont
0.68}}
    \psfrag{A}{\tiny{\fontfamily{phv}\selectfont
0.45}}
    \psfrag{z}{\tiny{\fontfamily{phv}\selectfont
0.23}}
    \psfrag{y}{\tiny{\fontfamily{phv}\selectfont
0}}
    \psfrag{F}{\tiny{\fontfamily{phv}\selectfont
\hspace{-0.3cm}805\hspace{1.5cm} 32}}
    \psfrag{E}{\tiny{\fontfamily{phv}\selectfont
\hspace{-0.38cm}803.8\hspace{1.37cm} 28.8}}
    \psfrag{D}{\tiny{\fontfamily{phv}\selectfont
\hspace{-0.38cm}802.5\hspace{1.37cm} 25.5}}
    \psfrag{K}{\tiny{\fontfamily{phv}\selectfont
\hspace{-0.38cm}801.3\hspace{1.37cm} 22.3}}
    \psfrag{H}{\tiny{\fontfamily{phv}\selectfont
\hspace{-0.29cm}800\hspace{1.51cm} 19}}
    \psfrag{G}{\tiny{\fontfamily{phv}\selectfont
0 \hspace{0.22cm} 1/3 \hspace {0.28cm} 2/3}}
    \psfrag{o}{\tiny{\fontfamily{phv}\selectfont
\hspace{-0.13cm}$\Sigma$ (-)}}
    \psfrag{M}{\tiny{\fontfamily{phv}\selectfont
\hspace{-0.7cm}frequency $f_\pm$ (Hz)}}
    \psfrag{L}{\tiny{\fontfamily{phv}\selectfont
\hspace{-0.3cm}$\beta_\pm$ (rad/s)}}
    \psfrag{2}{\tiny{\fontfamily{phv}\selectfont
\hspace{-0.25cm}16.1}}
    \psfrag{1}{\tiny{\fontfamily{phv}\selectfont
\hspace{-0.27cm}15.8}}
    \psfrag{Z}{\tiny{\fontfamily{phv}\selectfont
\hspace{-0.27cm}15.5}}
    \psfrag{Y}{\tiny{\fontfamily{phv}\selectfont
\hspace{-0.27cm}15.2}}
    \psfrag{X}{\tiny{\fontfamily{phv}\selectfont
\hspace{-0.25cm}14.9}}
   \psfrag{S}{\tiny{\fontfamily{phv}\selectfont
   7.02  \hspace{0.178cm}  7.33   \hspace{0.178cm} 7.64  \hspace{0.178cm} 7.96    \hspace{0.178cm} 8.27}}
      \psfrag{T}{\tiny{\fontfamily{phv}\selectfont
\hspace{-0.2cm}$V$ (m/s)}}
      \psfrag{U}{\tiny{\fontfamily{phv}\selectfont
\hspace{-0.2cm}$U$ (m/s)}}
      \psfrag{6}{\tiny{\fontfamily{phv}\selectfont
\hspace{-0.5cm}isolation (dB)}}
      \psfrag{R}{\tiny{\fontfamily{phv}\selectfont
10}}
      \psfrag{Q}{\tiny{\fontfamily{phv}\selectfont
-3.2}}
      \psfrag{P}{\tiny{\fontfamily{phv}\selectfont
-16}}
      \psfrag{O}{\tiny{\fontfamily{phv}\selectfont
-29}}
      \psfrag{N}{\tiny{\fontfamily{phv}\selectfont
-41}}
      \psfrag{7}{\textbf{\tiny{\fontfamily{phv}\selectfont
b\hspace{3.5cm}c}}}
      \psfrag{8}{}
      \psfrag{9}{\tiny{\fontfamily{phv}\selectfont
\hspace{-0.23cm}\textcolor{blue1}{$f_+$}\hspace{1.48cm}\textcolor{orangex}{$\beta_-$}}}
      \psfrag{0}{\tiny{\fontfamily{phv}\selectfont
\hspace{-0.22cm}\textcolor{orangex}{$f_-$}\hspace{1.47cm}\textcolor{blue1}{$\beta_+$}}}
         \psfrag{<}{\tiny{\fontfamily{phv}\selectfont
\hspace{-0.1cm}\textcolor{purple1}{$\gamma$}}}
          \psfrag{>}{\tiny{\fontfamily{phv}\selectfont
\hspace{0.18cm}stable\hspace{0.07cm} unstable}}
      \psfrag{+}{\tiny{\fontfamily{phv}\selectfont
\hspace{0.055cm}-1\hspace{0.15cm}1}}
      \psfrag{-}{\tiny{\fontfamily{phv}\selectfont
\hspace{-0.02cm}$\tilde{\Omega}_\theta$ [s\textsuperscript{-1}]}}
       \psfrag{?}{\tiny{\fontfamily{phv}\selectfont
\hspace{0.15cm}\textcolor{blue1}{co-swirl}}}
       \psfrag{*}{\tiny{\fontfamily{phv}\selectfont
\hspace{0.15cm}\textcolor{orangex}{counter-swirl}}}
    \psfrag{a}{\tiny{\fontfamily{phv}\selectfont
20.1\hspace{0.33cm}24.3}}
    \psfrag{q}{\tiny{\fontfamily{phv}\selectfont
\hspace{-0.05cm}$\overline{p}^2$ }}
\includegraphics[width=0.5\textwidth]{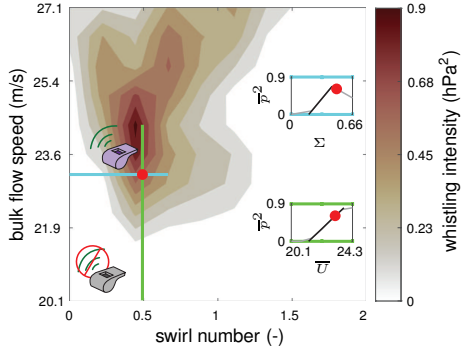}
\end{psfrags}
\caption{Bifurcation diagram of the circulator cavity. Shown is the whistling intensity, defined as the squared acoustic RMS amplitude $\overline{p}^2$, measured inside the cavity for different values of the bulk flow speed $\overline{U}=U+V$ and the swirl number $\Sigma=V/U$ (see Figure \ref{Appendix Figure 2} for a definition of $U$ and $V$). The operating point at $\overline{U}=23.1$ m/s and $\Sigma=0.49$ is marked by a red dot. The smaller insets show the variation of $\overline{p}^2$ (gray curves) along the cyan and green curves, respectively. The black lines in the smaller insets are linear regression lines to extrapolate the bifurcation points.}\label{Appendix Figure 3}
\end{figure}

The whistling phenomenon observed in the experiments can be described in low-order models, such as the ``$-$'' branch  in Eq. \eqref{modal dynamics 2m}, which, for $\vert s\rangle=0$ and $\beta_->\gamma$, describes a limit cycle arising from a supercritical Hopf bifurcation. Idealized theory predicts that the square of the modal amplitude $a_-$ varies linearly with the bifurcation parameter $\beta_-$ in the vicinity of the bifurcation point. The bifurcation is governed by the system nonlinearity shown in Fig. \ref{Appendix Figure 1}. In the experiments, we have two free parameters ($U$ and $V$), which results in a two-parameter Hopf bifurcation, as shown in  Fig. \ref{Appendix Figure 3}. To generate this inset, we recorded the acoustic pressure oscillations in the absence of \cc{waves incident to the cavity ports with} two microphones on the downstream wall of the cavity. \cc{These 10 s timetraces were recorded} over an equally spaced grid of values of the bulk velocity $\overline{U}$ and the swirl number $\Sigma=V/U$. For each measurement, we then computed the RMS amplitude of each of the two microphones and took the arithmetic \cc{mean} of these two values, which we denote \cc{by $\overline{p}$}. We used the average of the RMS amplitudes from two non-diametrically opposed microphones instead of the RMS of a single microphone to avoid the situation that the nodal line of the aeroacoustic mode (compare Fig. \ref{Figure 2}b) lies on the microphone during the measurement. The operating point ($\overline{p}^2=0.61$ hPa\textsuperscript{2}) for the scattering experiments, marked by the red dot, clearly lies in the \cc{limit cycle} regime. The smaller cyan and green insets show the variation of \cc{the whistling intensity} $\overline{p}^2$ (gray curves) along the cyan ($\overline{U}=23.1$m/s) and green ($\Sigma=0.49$) lines in Fig. \ref{Figure 3}c, respectively, and the black lines within the smaller insets are linear regressions to extrapolate the bifurcation points. In the cyan inset, the regression itself is trivial because the black line connects exactly two measuring points, but the estimation of the bifurcation point is nevertheless informative.

\subsection{Synchronization dynamics}
\label{Appendix_Sync}
\begin{figure}[h] % Figure 3
    \centering
    \psfrag{a}[c]{\hspace{20mm}\fontfamily{phv}\selectfont\footnotesize  1.7\hspace{1.5cm}1.8}
    \psfrag{b}[c]{\fontfamily{phv}\selectfont\footnotesize 1.7\hspace{2mm}1.8}
    \psfrag{c}[c]{\fontfamily{phv}\selectfont\footnotesize 1.9}
       \psfrag{3}[c]{\hspace{20mm}\fontfamily{phv}\selectfont\footnotesize  1.78\hspace{1.5cm}1.8}
    \psfrag{4}[c]{\fontfamily{phv}\selectfont\footnotesize 1.82}
    \psfrag{d}[c]{\fontfamily{phv}\selectfont\footnotesize 0}
    \psfrag{e}[c]{}
    \psfrag{f}[c]{\fontfamily{phv}\selectfont\footnotesize 100}
    \psfrag{g}[c]{}
    \psfrag{h}[c]{\fontfamily{phv}\selectfont\footnotesize 200}
    \psfrag{i}[c]{\hspace{-5mm} \fontfamily{phv}\selectfont\footnotesize 20}
    \psfrag{j}[c]{\hspace{-6mm} \fontfamily{phv}\selectfont\footnotesize 31.6}
    \psfrag{k}[c]{\hspace{-4.5mm} \fontfamily{phv}\selectfont\footnotesize 70}
    \psfrag{l}[c]{\hspace{-6mm} \fontfamily{phv}\selectfont\footnotesize 120}
    \psfrag{m}[c]{\hspace{-3mm}\fontfamily{phv}\selectfont\footnotesize 75}
    \psfrag{n}[c]{\hspace{-4mm}\fontfamily{phv}\selectfont\footnotesize 90}
    \psfrag{o}[c]{\hspace{-3mm}\fontfamily{phv}\selectfont\footnotesize 105}
    \psfrag{p}[c]{\hspace{-3mm}\fontfamily{phv}\selectfont\footnotesize 120}
    \psfrag{q}[c]{\hspace{-3mm}\fontfamily{phv}\selectfont\footnotesize 135}
    \psfrag{r}[c]{\hspace{-5mm}\fontfamily{phv}\selectfont\footnotesize -240}
    \psfrag{s}[c]{\hspace{-5mm}\fontfamily{phv}\selectfont\footnotesize -120}
    \psfrag{t}[c]{\hspace{-4mm}\fontfamily{phv}\selectfont\footnotesize 0}
    \psfrag{u}[c]{\hspace{-4mm}\fontfamily{phv}\selectfont\footnotesize 120}
    \psfrag{v}[c]{\hspace{-4mm}\fontfamily{phv}\selectfont\footnotesize 240}
    \psfrag{A}[c]{\footnotesize$f$ \fontfamily{phv}\selectfont\footnotesize (kHz)}
    \psfrag{B}[c]{\fontfamily{phv}\selectfont\footnotesize amp.\ $s_\mathrm{in}$\fontfamily{phv}\selectfont\footnotesize (Pa)}
    \psfrag{C}[c]{\footnotesize$f$ \fontfamily{phv}\selectfont\footnotesize (kHz)}
    \psfrag{D}[c]{\vspace{-3mm} \fontfamily{phv}\selectfont\footnotesize PSD (dB)}
    \psfrag{E}[c]{\hspace{0.5cm}\footnotesize$\Delta t$ \fontfamily{phv}\selectfont\footnotesize (ms)}
    \psfrag{F}[c]{$p$ \fontfamily{phv}\selectfont\footnotesize (Pa)}
       \psfrag{1}[c]{\hspace{0.1cm}\footnotesize$\Delta t$}
    \psfrag{2}[c]{\footnotesize$p$}
    \psfrag{Z}{\fontfamily{phv}\selectfont\footnotesize \textbf{a}}
    \psfrag{W}{\fontfamily{phv}\selectfont\footnotesize \textbf{b}}
    \psfrag{X}{\fontfamily{phv}\selectfont\footnotesize \textbf{c}}
    \psfrag{Y}{\fontfamily{phv}\selectfont\footnotesize \textbf{d}}
    \psfrag{K}[c]{\hspace{1cm}\fontfamily{phv}\selectfont\footnotesize async}
    \psfrag{L}[c]{\hspace{1cm}\fontfamily{phv}\selectfont\footnotesize sync}
    \includegraphics[width=0.65\textwidth]{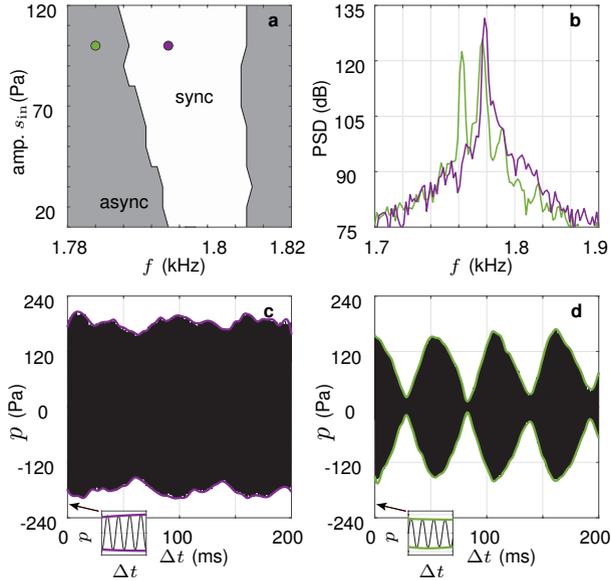}
    \caption{Synchronization dynamics of the nonlinear cavity mode. \textbf{a} Experimental Arnold tongue. Incident waves at fixed amplitudes  $s_{\text{in}}$ and frequencies $f$ were sent to the obstacle. By analyzing the relative peak height of the power spectrum density (PSD) of the forced mode, synchronized states can be identified. \textbf{b} A synchronized mode has an isolated peak at the forcing frequency. A non-synchronized mode has one peak at the self-sustained limit cycle frequency and another at the forcing frequency, as well as less dominant harmonics. \textbf{c} Pressure dynamics of a synchronized mode: quasi-sinusoidal oscillation at quasi-constant amplitude. \textbf{d} Non-synchronized states show beating oscillations due to interference of the self-sustained mode and the external forcing. The small insets in \textbf{c} and \textbf{d} show a couple of cycles of the same time traces.}
    \label{Appendix Figure 7}
\end{figure}

As described in the last section, we now consider a scenario where a limit cycle is reached at the conditions described above. Then, in addition to these self-sustained oscillations, using the compression driver mounted on the top arm, harmonic sound waves are sent through the channel towards the cavity. As we measure the acoustic pressure inside the cavity, at the same azimuthal position as the forcing waveguide, we can observe synchronization phenomena similar to those described in the book of Balanov et al. \cite{Balanov2009}. To analyze these dynamics, we performed repeated forced experiments at different incident wave amplitudes for fixed amplitude of the limit cycle. The results are presented in Fig. \ref{Appendix Figure 7}. Figure \ref{Appendix Figure 7}a illustrates the Arnold tongue \cite{Coombes19992086} representing the region in which synchronization of the incident waves takes place. As we can see, for the smallest incident wave amplitude, the synchronized region is almost a double its size at the largest forcing amplitude. This is consistent with the classic paradigm that synchronization requires large amplitudes and small detuning. For a stronger forcing, hence, larger detuning is acceptable to satisfy this double constraint, leading to a wider region in which synchronization can take place. We now take closer look at two specific points, marked by the purple and green dots in Fig. \ref{Appendix Figure 7}a at a forcing amplitude of ${s_{\text{in}}}= 100$ Pa, we can visually compare the synchronized and asynchronous modes. For the latter, as we can see from Fig. \ref{Appendix Figure 7}b, there are two dominant peaks in the PSD while for the synchronized we observe just one. The non-synchronized mode has a peak at the self-sustained limit cycle around 800 Hz, another at the forcing frequency and a significantly less dominant harmonic peak at around 820 Hz. For the sychronized state, a peak is only observed at the forcing frequency. Figure \ref{Appendix Figure 7}c shows the pressure dynamics for a synchronized mode, which can be compared to those of an asynchronous mode as shown in Fig. \ref{Appendix Figure 7}d. While the synchronized mode shows roughly constant amplitude, significant beating oscillations accompany the asynchronous state, stemming from the interplay between the coexisting, detuned peaks of the forced and the self-sustained oscillations.

\subsection{Zeeman bias modeling}
\label{Appendix_Zeeman}
\begin{figure}[h]%
\centering
\begin{psfrags}
    \psfrag{a}{\tiny{\fontfamily{phv}\selectfont
\hspace{-0.02cm}min}}
    \psfrag{c}{\tiny{\fontfamily{phv}\selectfont
\hspace{-0.02cm}max}}
    \psfrag{b}{\tiny{\fontfamily{phv}\selectfont
\hspace{-0.02cm}0}}
    \psfrag{d}{\tiny{\fontfamily{phv}\selectfont
\hspace{-0.5cm}ac. pressure}}
    \psfrag{e}{\tiny{\fontfamily{phv}\selectfont
0 \hspace{0.39cm} 0.5 \hspace{0.39cm} 1
 \hspace{0.39cm} 1.5 \hspace{0.39cm} 2}}
    \psfrag{n}{\tiny{\fontfamily{phv}\selectfont
\hspace{-0.25cm}27.1}}
    \psfrag{m}{\tiny{\fontfamily{phv}\selectfont
\hspace{-0.25cm}25.4}}
    \psfrag{l}{\tiny{\fontfamily{phv}\selectfont
\hspace{-0.35cm}23.6}}
    \psfrag{k}{\tiny{\fontfamily{phv}\selectfont
\hspace{-0.3cm}21.9}}
    \psfrag{j}{\tiny{\fontfamily{phv}\selectfont
\hspace{-0.3cm}20.1}}
    \psfrag{p}{\tiny{\fontfamily{phv}\selectfont
\hspace{-0.75cm}bulk flow speed $\overline{U}$ (m/s)}}
    \psfrag{I}{\tiny{\fontfamily{phv}\selectfont
\hspace{-0.6cm}swirl number $\Sigma$ (-)}}
    \psfrag{4}{\tiny{\fontfamily{phv}\selectfont
\hspace{-0.05cm}$\Sigma$}}
    \psfrag{q}{\tiny{\fontfamily{phv}\selectfont
\hspace{-0.15cm}$\overline{p}^2$ }}
    \psfrag{3}{\tiny{\fontfamily{phv}\selectfont
\hspace{-1.2cm}whistling intensity $\overline{p}^2$ (hPa\textsuperscript{2})}}
    \psfrag{t}{\tiny{\fontfamily{phv}\selectfont
0\hspace{0.2cm}0.9}}
    \psfrag{t}{\tiny{\fontfamily{phv}\selectfont
0\hspace{0.2cm}0.9}}
    \psfrag{w}{\tiny{\fontfamily{phv}\selectfont
$\overline{U}$}}
    \psfrag{j}{\tiny{\fontfamily{phv}\selectfont
\hspace{-0.28cm}20.1}}
    \psfrag{u}{\tiny{\fontfamily{phv}\selectfont
\hspace{-0.28cm}24.3}}
    \psfrag{f}{\tiny{\fontfamily{phv}\selectfont
\hspace{-0.1cm}0}}
    \psfrag{x}{\tiny{\fontfamily{phv}\selectfont
\hspace{-0.2cm}0.9}}
    \psfrag{5}{\tiny{\fontfamily{phv}\selectfont
0\hspace{0.35cm}2/3}}
    \psfrag{C}{\tiny{\fontfamily{phv}\selectfont
0.9}}
    \psfrag{B}{\tiny{\fontfamily{phv}\selectfont
0.68}}
    \psfrag{A}{\tiny{\fontfamily{phv}\selectfont
0.45}}
    \psfrag{z}{\tiny{\fontfamily{phv}\selectfont
0.23}}
    \psfrag{y}{\tiny{\fontfamily{phv}\selectfont
0}}
    \psfrag{F}{\tiny{\fontfamily{phv}\selectfont
\hspace{-0.3cm}805\hspace{1.63cm} 32}}
    \psfrag{E}{\tiny{\fontfamily{phv}\selectfont
\hspace{-0.38cm}803.8\hspace{1.5cm} 28.8}}
    \psfrag{D}{\tiny{\fontfamily{phv}\selectfont
\hspace{-0.38cm}802.5\hspace{1.5cm} 25.5}}
    \psfrag{K}{\tiny{\fontfamily{phv}\selectfont
\hspace{-0.38cm}801.3\hspace{1.5cm} 22.3}}
    \psfrag{H}{\tiny{\fontfamily{phv}\selectfont
\hspace{-0.29cm}800\hspace{1.64cm} 19}}
    \psfrag{G}{\tiny{\fontfamily{phv}\selectfont
0 \hspace{0.23cm} 1/3 \hspace {0.23cm} 2/3}}
    \psfrag{o}{\tiny{\fontfamily{phv}\selectfont
\hspace{-0.1cm}$\Sigma$ (-)}}
    \psfrag{M}{\tiny{\fontfamily{phv}\selectfont
\hspace{-0.3cm} $f_\pm$ (Hz)}}
    \psfrag{L}{\tiny{\fontfamily{phv}\selectfont
\hspace{-0.3cm}$\beta_\pm$ (rad/s)}}
    \psfrag{2}{\tiny{\fontfamily{phv}\selectfont
\hspace{-0.25cm}16.1}}
    \psfrag{1}{\tiny{\fontfamily{phv}\selectfont
\hspace{-0.27cm}15.8}}
    \psfrag{Z}{\tiny{\fontfamily{phv}\selectfont
\hspace{-0.27cm}15.5}}
    \psfrag{Y}{\tiny{\fontfamily{phv}\selectfont
\hspace{-0.27cm}15.2}}
    \psfrag{X}{\tiny{\fontfamily{phv}\selectfont
\hspace{-0.25cm}14.9}}
   \psfrag{S}{\tiny{\fontfamily{phv}\selectfont
   7.02  \hspace{0.178cm}  7.33   \hspace{0.178cm} 7.64  \hspace{0.178cm} 7.96    \hspace{0.178cm} 8.27}}
      \psfrag{T}{\tiny{\fontfamily{phv}\selectfont
\hspace{-0.2cm}$V$ (m/s)}}
      \psfrag{U}{\tiny{\fontfamily{phv}\selectfont
\hspace{-0.2cm}$U$ (m/s)}}
      \psfrag{6}{\tiny{\fontfamily{phv}\selectfont
\hspace{-0.5cm}isolation (dB)}}
      \psfrag{R}{\tiny{\fontfamily{phv}\selectfont
10}}
      \psfrag{Q}{\tiny{\fontfamily{phv}\selectfont
-3.2}}
      \psfrag{P}{\tiny{\fontfamily{phv}\selectfont
-16}}
      \psfrag{O}{\tiny{\fontfamily{phv}\selectfont
-29}}
      \psfrag{N}{\tiny{\fontfamily{phv}\selectfont
-41}}
      \psfrag{7}{\textbf{\tiny{\fontfamily{phv}\selectfont
b\hspace{3.5cm}c}}}
      \psfrag{8}{}
      \psfrag{9}{\tiny{\fontfamily{phv}\selectfont
\hspace{-0.23cm}\textcolor{blue1}{$f_+$}\hspace{1.61cm}\textcolor{orangex}{$\beta_-$}}}
      \psfrag{0}{\tiny{\fontfamily{phv}\selectfont
\hspace{-0.22cm}\textcolor{orangex}{$f_-$}\hspace{1.6cm}\textcolor{blue1}{$\beta_+$}}}
         \psfrag{<}{\tiny{\fontfamily{phv}\selectfont
\hspace{-0.1cm}\textcolor{purple1}{$\gamma$}}}
          \psfrag{>}{\tiny{\fontfamily{phv}\selectfont
\hspace{0.18cm}stable\hspace{0.15cm} limit cycle}}
      \psfrag{+}{\tiny{\fontfamily{phv}\selectfont
\hspace{0.055cm}-1\hspace{0.15cm}1}}
      \psfrag{-}{\tiny{\fontfamily{phv}\selectfont
\hspace{-0.02cm}$\tilde{\Omega}_\theta$ [s\textsuperscript{-1}]}}
       \psfrag{?}{\tiny{\fontfamily{phv}\selectfont
\hspace{0.15cm}\textcolor{blue1}{co-swirl}}}
       \psfrag{*}{\tiny{\fontfamily{phv}\selectfont
\hspace{0.15cm}\textcolor{orangex}{counter-swirl}}}
\includegraphics[width=0.35\textwidth]{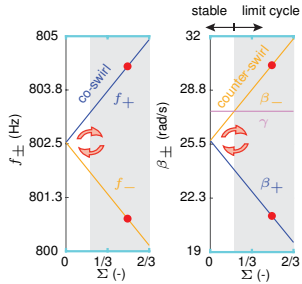}
\end{psfrags}
\caption{Modeled variation of the eigenfrequencies $f_\pm$, the linear gains $\beta_\pm$ and the decay rate $\gamma$ with respect to the swirl number $\Sigma$. The bias introduced by the swirl governs the Zeeman splitting of the eigenfrequencies that are degenerate at $\Sigma=0$. The values at the operating point are marked by red dots. The ``$+$'' and ``$-$'' branches correspond to the modes spinning with and against the swirling flow, respectively.} \label{Appendix Figure 4}
\end{figure}

As in \cite{Fleury2014516}, the broken symmetry of the theoretical model, expressed here by $\omega_+\neq \omega_-$ and $\beta_+\neq\beta_-$, corresponds in the experiments to the bias introduced by an azimuthal flow. Under the simplifying assumption that $f_\pm=\omega_\pm/2\pi$ and $\beta_\pm$ depend linearly on $\Sigma\neq 0$ and that for $\Sigma=0$, the system is perfectly symmetric, we obtain the variations of these parameters over the swirl number $\Sigma$ (see Figure \ref{Appendix Figure 4}). In this case, $a_+$ corresponds to the mode spinning \textit{with} the swirling flow and $a_-$ to the mode spinning against it. Note that $f_\pm$ and $\beta_\pm$ split in opposite direction in our model, which is consistent with experiments and numerical simulations performed on the same cavity and presented in \cite[Fig. 5]{Abel2023_2}. The bifurcation points extrapolated from the experiments and from the theoretical model both lie at $\Sigma=0.19$ with less than 1$\%$ relative deviation from each other, showing that the model and the experiments are consistent in their representation of the limit cycle's stability boundary.

\subsection{Transmission without and with flow}
\label{Appendix_Unbiased}
\begin{figure}[h]%
\centering
\begin{psfrags}
    \psfrag{d}{\tiny{\fontfamily{phv}\selectfont
\hspace{-0.05cm}780\hspace{0.28cm}820}}
    \psfrag{a}{\tiny{\fontfamily{phv}\selectfont
\hspace{-0.12cm}0}}
    \psfrag{b}{\tiny{\fontfamily{phv}\selectfont
\hspace{-0.22cm}0.5}}
    \psfrag{c}{\tiny{\fontfamily{phv}\selectfont
\hspace{-0.12cm}1}}
    \psfrag{f}{\tiny{\fontfamily{phv}\selectfont
\hspace{-0.6cm}frequency (Hz)}}
    \psfrag{e}{\tiny{\fontfamily{phv}\selectfont
\hspace{-0.5cm}transmission}}
    \psfrag{l}{\tiny{\fontfamily{phv}\selectfont
$ \Delta s $ (-)}}
    \psfrag{g}{\tiny{\fontfamily{phv}\selectfont
\hspace{-0.3cm}0\hspace{0.73cm}1\hspace{0.73cm}2\hspace{0.75cm}3\hspace{0.74cm}4\hspace{0.73cm}5\hspace{0.73cm}6\hspace{0.73cm}7\hspace{0.73cm}8\hspace{0.73cm}9\hspace{0.645cm}10\hspace{0.645cm}11}}
\includegraphics[width=0.88\textwidth]{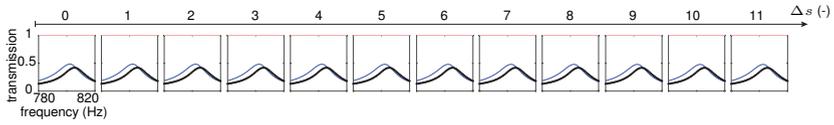}
\end{psfrags}
\caption{Measured transmission coefficients in the no-flow case for different values of the incident wave amplitude $\Delta s$ (see Sec. \ref{Section 4}) The black dotted curves correspond to $T_{1\shortrightarrow 3}$, the solid blue curves correspond to $T_{1\shortrightarrow 2}$ and the red lines correspond to 100$\%$ transmission.}\label{Appendix Figure 5}
\end{figure}

\begin{figure}[h]%
\centering
\begin{psfrags}
    \psfrag{a}{\tiny{\fontfamily{phv}\selectfont
780\hspace{0.4cm}820}}
    \psfrag{d}{\tiny{\fontfamily{phv}\selectfont
\hspace{-0.05cm}0}}
    \psfrag{e}{\tiny{\fontfamily{phv}\selectfont
\hspace{-0.05cm}1}}
    \psfrag{f}{\tiny{\fontfamily{phv}\selectfont
\hspace{-0.07cm}2}}
    \psfrag{i}{\tiny{\fontfamily{phv}\selectfont
\hspace{-0.6cm}frequency (Hz)}}
    \psfrag{j}{\tiny{\fontfamily{phv}\selectfont
\hspace{-0.5cm}transmission}}
    \psfrag{l}{\tiny{\fontfamily{phv}\selectfont
$V$ (m/s)}}
    \psfrag{k}{\tiny{\fontfamily{phv}\selectfont
$U$ (m/s)}}
    \psfrag{h}{\tiny{\fontfamily{phv}\selectfont
\hspace{-0cm}14.9\hspace{0.69cm}15.1\hspace{0.69cm}15.3\hspace{0.69cm}
15.5\hspace{0.69cm}15.7\hspace{0.69cm}15.9\hspace{0.69cm}16.1}}
    \psfrag{g}{\tiny{\fontfamily{phv}\selectfont
\hspace{0.07cm}7.02\hspace{0.71cm}7.23\hspace{0.71cm}7.43\hspace{0.71cm}
7.64\hspace{0.71cm}7.85\hspace{0.71cm}8.06\hspace{0.71cm}8.27}}
\includegraphics[width=0.7\textwidth]{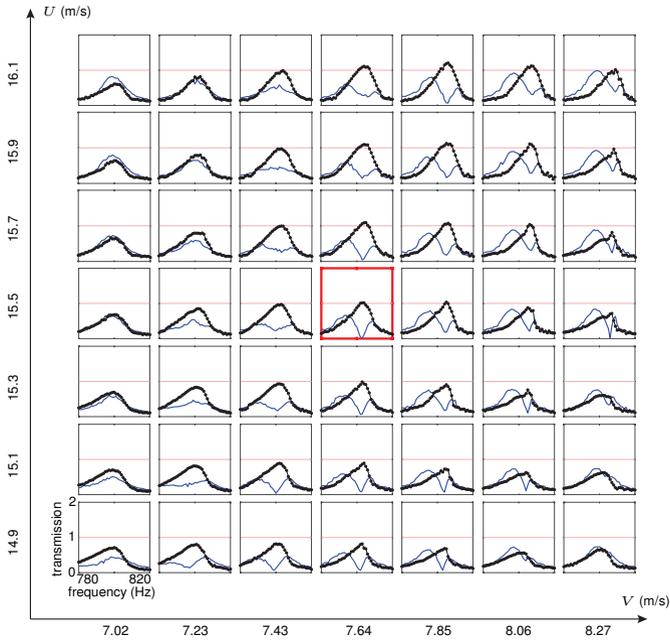}
\end{psfrags}
\caption{Sensitivity of the measured amplitude transmission coefficients at $\Delta s=9$ over different values of the axial ($U$) and azimuthal ($V$) flow components. The black dotted curves correspond to $T_{1\shortrightarrow 3}$, the solid blue curves correspond to $T_{1\shortrightarrow 2}$ and the red lines correspond to 100$\%$ transmission. The operating point is marked with a thick red box around the respective inset.}\label{Appendix Figure 6}
\end{figure}

If no flow is imposed on the cavity, then it behaves like a reciprocal, lossy and linear 3-port scatterer, as shown in  Fig. \ref{Appendix Figure 5}. This shows that, besides the whistling tone induced by the air flow, there are no non-linear effects present in the system.
If a flow is imposed on the cavity, a non-reciprocal lossless transmission can be achieved, as shown in Fig. \ref{Appendix Figure 6}.

\newpage

~\\

\clearpage

~\\
~\\
\begin{center}
  \Large Supplementary notes to \\ ``Synchronization-based lossless \\ non-reciprocal scattering''\\[1cm]
 \large  Tiemo Pedergnana,$^{1}$ Abel Faure-Beaulieu$^{1}$, Romain Fleury$^{2*}$ \\ and Nicolas Noiray$^{1*}$\\[.5cm]
  { ${}^1$Department of Mechanical and Process Engineering, ETH \\ Z\"urich, Z\"urich, Switzerland\\
  ${}^2$Institute of Electrical and Micro Engineering, EPFL, Vaud, Switzerland}\\[.5cm]
  ${}^*$Corresponding authors. E-mails: \textcolor{blue}{romain.fleury@epfl.ch};\\ \textcolor{blue}{noirayn@ethz.ch};
\\[1cm]
\end{center}
\thispagestyle{empty}

\setcounter{equation}{0}
\setcounter{figure}{0}
\setcounter{table}{0}
\setcounter{page}{1}
\renewcommand{\theequation}{S\arabic{equation}}
\renewcommand{\thefigure}{S\arabic{figure}}

\clearpage

\section*{Theory}
\subsection*{Scattering matrix derivation}
To model the non-linear scattering process, we propose a coupled-mode theory for the cavity modes $a=(a_+,a_-)^T$, which are forced by the incoming wave $\vert s_\mathrm{in}\rangle$, leading to an outgoing wave $\vert s_\mathrm{out}\rangle$. The following equations form the theory presented in this work:\footnote{For generality, we derive the scattering matrix for the general case where $\gamma_+$ and $\kappa_+$ are not necessarily equal to $\gamma_-$ and $\kappa_-$, respectively. In the main text, only the symmetric case with $\gamma_\pm=\gamma$, $\kappa_\pm=\kappa$ is considered. }
\begin{eqnarray}
    \dot{a}_\pm&=&i\omega_{_\pm}a_\pm  -\gamma_\pm a_\pm  +\frac{\beta_\pm a_\pm }{1+\kappa_\pm \vert a_\pm\vert^2}  +D^\dag_\pm \vert s_{\mathrm{in}}\rangle,\label{modal dynamics 2}\\
   \vert s_{\mathrm{out}}\rangle&=&C \vert s_{\mathrm{in}}\rangle + D a, \label{input-output 2}
\end{eqnarray}
where a dot denotes the time derivative, $\omega_{\pm}$ are the eigenfrequencies, $\gamma_\pm$ are the decay rates \cc{corresponding to dissipative processes in the cavity and radiation losses}, $\beta_\pm\in\mathbb{R}^+$ are the linear gain coefficients, $\kappa_\pm\in\mathbb{R}^+$ are the non-linearity constants, $C\in \mathbb{R}^{3\times 3}$ is the \cc{background scattering matrix} and the matrix $D=[D_+,D_-]\in\mathbb{C}^{3\times 2}$ with column vectors $D_\pm$ describes the wave-mode coupling. When the system is harmonically excited from the $j^\mathrm{th}$ port, we have $\vert s_\mathrm{in}\rangle=s e^{i\omega t} \vert j\rangle$, where $s=\sqrt{\langle s_\mathrm{in} \vert  s_\mathrm{in} \rangle}\in\mathbb{R}^+$ is the incident wave amplitude and $\vert j \rangle$ is the unit vector in $j$-direction. The vector $a=(a_+,a_-)^T\in\mathbb{C}^2$ in Eq. \eqref{input-output 2}, where $(\cdot)^T$ is the transpose, comprises both cavity modes. 

For simplicity, to compute the forced response, we write $\vert a_\pm \vert=A_\pm$ and we assume that the modes are perfectly synchronized with the excitation: $a_\pm=A_\pm e^{i(\omega t+\varphi_\pm)}$, where $\vert\cdot\vert\in \mathbb{R}^+$ is the absolute value and $A_\pm$ and $\varphi_\pm$ are the amplitudes and phases of the forced response, respectively. With this assumption, the modal dynamics \eqref{modal dynamics 2} become
\begin{eqnarray}
  \big[\dot{A}_\pm+A_\pm i(\omega+\dot{\varphi}_\pm)\big]&=&\nonumber \\
   &&\hspace{-2cm}\Big(i\omega_{0,_\pm} -\gamma_\pm +\frac{\beta_\pm}{1+\kappa_\pm A_\pm^2} \Big)A_\pm +\vert D_{\pm,j} \vert s e^{-i(\arg D_{\pm,j}+\varphi_\pm)}.
\end{eqnarray}
Separating real and imaginary parts yields
\begin{eqnarray}
\dot{A}_\pm&=&-\gamma_\pm A_\pm+\frac{\beta_\pm A_\pm}{1+\kappa_\pm A_\pm^2}+\vert D_{\pm,j} \vert s \cos{(\arg D_{\pm,j}+\varphi_\pm)}, \\
    \dot{\varphi}_\pm&=&\Delta_\pm-\frac{\vert D_{\pm,j} \vert s\sin{(\arg D_{\pm,j}+\varphi_\pm)}}{ A_\pm},
\end{eqnarray}
where the terms $\Delta_\pm=\omega_{\pm}-\omega$ represent the detuning between the forcing and the respective oscillators. The unforced ($s=0$) limit cycles $A_\pm=A_{0,\pm}\in\mathbb{R}^+$, $\varphi_\pm=\varphi_{0,\pm}\in[0, 2\pi)$ are given by 
\begin{eqnarray}
A_{0,\pm}&=&\sqrt{\frac{1}{\kappa_\pm}\Big(\frac{\beta_\pm}{\gamma_\pm}-1\Big)}, \label{Limit cycle formula}
\end{eqnarray}
and can only exist when the respective \cc{linear gain} is greater than the dissipation: $\beta_\pm>\gamma_\pm$. Assuming this is the case, the linearized unforced modal dynamics around the limit cycles $A_{0,\pm}$,
\begin{eqnarray}
  \dot{A}'_\pm&=&-\Bigg(\gamma_\pm+\frac{\beta_\pm (\kappa_\pm A_{0,\pm}^2-1)}{(1+\kappa_\pm A_{0,\pm}^2)^2}\Bigg)A'_\pm\nonumber\\
  &=&-\Bigg(\dfrac{2\gamma_\pm(\beta_\pm-\gamma_\pm)}{\beta_\pm}\Bigg)A'_\pm
\end{eqnarray}
are asymptotically stable. If $\beta_\pm<\gamma_\pm$, the limit cycles $A_{0,\pm}$ given by Eq. \eqref{Limit cycle formula} do not exist and the only stationary solution \cc{is} the trivial one: $A_\pm\equiv 0$. In that case, the linearized dynamics around the trivial solution,
\begin{eqnarray}
  \dot{A}'_\pm&=&-(\gamma_\pm-\beta_\pm)A'_\pm, \label{Trivial solution stability}
\end{eqnarray}
are also asymptotically stable. If there does exist a limit cycle, then we see from Eq. \eqref{Trivial solution stability} that the trivial solution is linearly unstable. Since the unforced modal equations \eqref{modal dynamics 2} are decoupled, depending on the sign of $\beta_\pm-\gamma_\pm$, each of the modes $a_\pm$ has exactly one linearly stable stationary solution, which can be either a limit cycle or the trivial solution. 

Having discussed the modes' unforced dynamics, we now compute their forced response by setting $\dot{A}_\pm=\dot{\varphi}_\pm=0$, leading to the following expression:
\begin{eqnarray}
    \bigg[\Big(\vert D_{\pm,j} \vert s\Big)^2-(\Delta_\pm^2+\gamma_\pm^2)A_\pm^2\bigg](1+\kappa_\pm A_\pm^2)^2+2\gamma_\pm \beta_\pm A_\pm^2(1+\kappa_\pm A_\pm^2)\nonumber\\
   -\beta_\pm^2 A_\pm^2=0, 
\end{eqnarray}
which can be written as two cubic equations for $A_\pm^2$:
\begin{eqnarray}
    0&=&\alpha_{0,\pm}+\alpha_{1,\pm} A_\pm^2 +\alpha_{2,\pm} A_\pm^4 +\alpha_{3,\pm} A_\pm^6, \label{forced response A}
 \end{eqnarray}
    where
\begin{eqnarray}
    \alpha_{0,\pm}&=&\vert  D_{\pm,j} \vert^2 s^2\nonumber\\
    \alpha_{1,\pm}&=&2 \vert D_{\pm,j} \vert^2 s^2 \kappa_\pm-(\Delta_\pm^2+\gamma_\pm^2)+\beta_\pm(2\gamma_\pm-\beta_\pm)\nonumber\\
    \alpha_{2,\pm}&=&\kappa_\pm^2\vert D_{\pm,j} \vert^2 s^2-2\kappa_\pm(\Delta_\pm^2+{\gamma_\pm^2})+2\gamma_\pm\beta_\pm\kappa_\pm\nonumber\\
    \alpha_{3,\pm}&=&-\kappa_\pm^2(\Delta_\pm^2+{\gamma_\pm^2})\nonumber
\end{eqnarray}
The linearly stable branches of $\varphi_\pm$ of the forced response are given by 
\begin{eqnarray}
    \varphi_\pm=-\arg D_{\pm,j} +\arcsin\Big(\frac{\Delta_\pm A_\pm}{\vert D_{\pm,j} \vert s}\Big). \label{forced response phi}
\end{eqnarray}
At infinitely large amplitudes $A_\pm$, the term $-\gamma_\pm A_\pm$ dominates the $\beta_\pm$-modulated term in the equations for $\dot{A}_\pm$. In this limit, therefore, the dynamics \eqref{modal dynamics 2} converge to
\begin{eqnarray}
    \dot{a}_\pm&=&\Big(i\omega_\pm -\gamma_\pm  \Big)a_\pm +D^\dag \vert s_{\mathrm{in}}\rangle,\label{modal dynamics 2, large A}\\
   \vert s_{\mathrm{out}}\rangle&=&C \vert s_{\mathrm{in}}\rangle + D a, \label{input-output 2, large A}
\end{eqnarray}
which are exactly equivalent to the classic temporal coupled-mode theory\footnote{Fan, S., Suh, W. \& Joannopoulos, J. Temporal coupled-mode theory for the fano resonance in optical resonators. \textit{J. Opt. Soc. Am. A} \textbf{20} (3), 569–572 (2003). \url{https://doi.org/10.1364/JOSAA.20.000569};\\ Wang, Z. \& Fan, S. Magneto-optical defects in two-dimensional photonic crystals. \textit{Appl. Phys. B} \textbf{81}, 369–375 (2005). \url{https://doi.org/10.1007/ s00340-005-1846-x}}. To model this limiting case in a manner which is consistent with non-reciprocal scattering, following the method from Pedergnana and Noiray\footnote{Pedergnana, T. \& Noiray, N. Superradiant scattering by a limit cycle. \textit{Phys. Rev. Appl.} \textbf{20}, 034068 (2023). \url{https://doi.org/10.1103/PhysRevApplied. 20.034068}.}, we superimpose a non-reciprocal optimal scattering matrix $S_*$ on a purely reflectional background $C$:
\begin{eqnarray}
     S_*=-\begin{pmatrix}
0 & 1 & 0\\
0 & 0 & 1\\
1 & 0 & 0
\end{pmatrix},\quad
    C=-\begin{pmatrix}
1 & 0 & 0\\
0 & 1 & 0\\
0 & 0 & 1
\end{pmatrix}.
\end{eqnarray}
The matrix $\sigma S_*-C$, where $\sigma\in[0,1]$ is the unitarity factor, has the following eigenvalues $\mu_i$ and corresponding eigenvectors $v_i$, $i=1,2,3$:
\begin{align}
\mu_1&=\sigma-1,\quad &v_1&=\frac{1}{\sqrt{3}}\begin{pmatrix}
1 \\
1\\
1
\end{pmatrix},\\
\mu_2&=\sigma e^{\frac{i\pi }{3}}+1,\quad &v_2&=\frac{1}{\sqrt{3}}\begin{pmatrix}
e^{-\frac{2\pi i}{3}}\\
e^{\frac{2\pi i}{3}}\\
1
\end{pmatrix},\\
\mu_3&=\sigma e^{\frac{-i\pi }{3}}+1=\mu_2^*,\quad &v_3&=\frac{1}{\sqrt{3}}\begin{pmatrix}
e^{\frac{2\pi i}{3}}\\
e^{\frac{-2\pi i}{3}}\\
1
\end{pmatrix}=v_2^*,
\end{align}
where $(\cdot)^*$ is the complex conjugate. Formally approximating the linear part of the scattering matrix by $\sigma S_*$ yields a low-rank approximation problem for $\sigma S_*-C$ by the coupling matrix $D$:
\begin{eqnarray}
     \sigma S_*-C\approx D F^{-1} D^\dag, \label{Low-rank approximation}
\end{eqnarray}
where
\begin{eqnarray}
    F=\begin{pmatrix}
    i(\omega-\omega_+)+\gamma_+& 0\\
    0 & i(\omega-\omega_-)+\gamma_-
    \end{pmatrix}. \label{Definition F}
\end{eqnarray}
In the recent work from Pedergnana and Noiray\footnote{Pedergnana, T. \& Noiray, N. Superradiant scattering by a limit cycle. \textit{Phys. Rev. Appl.} \textbf{20}, 034068 (2023). \url{https://doi.org/10.1103/PhysRevApplied. 20.034068}.}, consistency conditions are given to determine $D$ and perfect matching condition for a scatterer with one mode and two ports. The latter condition describes an optimal parameter configuration at which the dominant eigenspace of $\sigma S_*-C$ is best approximated by $D F^{-1} D^\dag$. The goal of the consistency conditions is to ensure that the linear scattering matrix derived in this section successfully approximates the dominant sub-spectrum of $\sigma S_*$. Because $D$ is the only unknown in our formulation of the wave-mode coupling, we consider the problem \eqref{Low-rank approximation} instead of directly approximating $\sigma S_*$ by $C+D F^{-1} D^\dag$. We formulate the following requirements for a scatterer with two modes and three ports:
\vspace{0.5cm}
\begin{itemize}
    \item[(I)] The singular (rank $2$) matrix $DF^{-1}D^\dag$ spans the subspace corresponding to the eigenvalues of $\sigma S_*-C$ with the largest absolute values $\vert \mu_{i} \vert$, $i=1,2,3$. In general, $DF^{-1}D^\dag$ does not share any eigenvalues with $\sigma S_*-C$.\vspace{0.5cm}
    \item[(II)] At perfect matching for a non-unitary process $\sigma<1$, the two out of three eigenvalues with largest absolute values and the corresponding eigenvectors of $DF^{-1}D^\dag$ coincide with the two of $\sigma S_*-C$ with largest absolute value. 
\end{itemize}\vspace{0.5cm}
We show that (R1) and (R2) imply that at perfect matching, $\sigma S_*$ and $C+DF^{-1}D^\dag$ also share the sub-spectrum corresponding to the two eigenvalues with largest absolute value. To understand how $D$ is determined by (R1) and (R2), we write out the right-hand-side of the approximation \eqref{Low-rank approximation} explicitly:
\begin{eqnarray}
     DF^{-1}D^\dag&=&\begin{pmatrix}
D_+ & D_-
\end{pmatrix}\begin{pmatrix}
\dfrac{1}{g_+ \gamma_+} & 0\\
0 & \dfrac{1}{g_- \gamma_-}
\end{pmatrix}\begin{pmatrix}
D_+^\dag \\
D_-^\dag
\end{pmatrix}\\
&=&\dfrac{D_+ D_+^\dag}{g_+\gamma_+}+\dfrac{D_- D_-^\dag}{g_-\gamma_-}, \label{RHS of approximation}
\end{eqnarray}
where we have defined $g_\pm \gamma_\pm=i(\omega-\omega_\pm)+\gamma_\pm$. In comparison, the left-hand-side of the approximation \eqref{Low-rank approximation} can be expressed as 
\begin{eqnarray}
     \sigma S_*-C=v_1 v_1^\dag \mu_1+v_2 v_2^\dag \mu_2+v_3 v_3^\dag \mu_3. \label{relative s matrix eigendecomposition}
\end{eqnarray}
To approximate the dominant part of the spectrum of $\sigma S_*-C$, noting that $\vert\mu_{2}\vert=\vert\mu_{3}\vert>\vert\mu_1\vert$, we make the Ansatz 
\begin{eqnarray}
     D_+=\sqrt{\gamma_+\alpha(\sigma)}\, v_{2},\quad  D_-=\sqrt{\gamma_-\alpha(\sigma)}\, v_{3}, \label{D ansatz}
\end{eqnarray}
where $\alpha\in\mathbb{R}^+$. In the unitary case ($\sigma=1$), $\alpha=2$ and the perfect matching condition is $\omega_\pm=\omega\pm\gamma_\pm/\sqrt{3}$ \footnote{Wang, Z. \& Fan, S. Magneto-optical defects in two-dimensional photonic crystals. \textit{Appl. Phys. B} \textbf{81}, 369–375 (2005). \url{https://doi.org/10.1007/ s00340-005-1846-x}}. By analogy, we assume the perfect matching condition for $\sigma<1$ has the form
\begin{eqnarray}
     \omega_\pm=\omega\pm\chi(\sigma)\gamma_\pm, \label{perfect matching ansatz}
\end{eqnarray}
where $\chi\in\mathbb{R}^+$. Substituting the assumed relations \eqref{D ansatz} and \eqref{perfect matching ansatz} into Eq. \eqref{RHS of approximation} yields, at perfect matching,
\begin{eqnarray}
     DF^{-1}D^\dag&=&\alpha(\sigma)\bigg[\dfrac{v_2 v_2^\dag}{1-i\chi(\sigma)}+\dfrac{v_3 v_3^\dag}{1+i\chi(\sigma)}\bigg]. \label{RHS of approximation after substitution}
\end{eqnarray}
Comparing Eq. \eqref{RHS of approximation after substitution} to Eq. \eqref{relative s matrix eigendecomposition}, we require that
\begin{eqnarray}
     \dfrac{\alpha(\sigma)}{1-i\chi(\sigma)}=\mu_2\quad\leftrightarrow\quad \dfrac{\alpha(\sigma)}{1+i\chi(\sigma)}=\mu_3. \label{int. res. 1}
\end{eqnarray}
Since $\alpha\in\mathbb{R}$, $\mu_{2,3}=1+\sigma(1\pm i\sqrt{3})/2$ and $\vert \mu_{2,3}\vert^2 =\sigma^2+\sigma+1$, Eq. \eqref{int. res. 1} has the following unique solutions for $\alpha$ and $\chi$:
\begin{eqnarray}
     \alpha&=&\dfrac{2(\sigma^2+\sigma+1)}{\sigma+2},\\
     \chi&=&\frac{\sqrt{3}\sigma}{\sigma+2},
\end{eqnarray}
which, for $\sigma= 1$, reduce to the well-known values $\alpha=2$ and $\chi=1/\sqrt{3}$ found in literature on unitary scattering. Using the above results, the coupling matrix $D$ becomes
\begin{eqnarray}
     D=\sqrt{\dfrac{2(\sigma^2+\sigma+1)}{3(\sigma+2)}}\begin{pmatrix}
\sqrt{\gamma_+} e^{-\frac{2\pi i}{3}} &\sqrt{\gamma_-}e^{\frac{2\pi i}{3}} \\
\sqrt{\gamma_+} e^{\frac{2\pi i}{3}}&\sqrt{\gamma_-}e^{-\frac{2\pi i}{3}}  \\
\sqrt{\gamma_+}&\sqrt{\gamma_-}
\end{pmatrix}. \label{coupling matrix D}
\end{eqnarray}
By computing the linear scattering matrix $C+DF^{-1}D^\dag$, it is straightforward to verify that for a perfectly matched sub-unitary scatterer ($\sigma<1$), its two eigenvalues with largest absolute value coincide with those of $\sigma S_*$. Indeed, at perfect matching,
\begin{eqnarray}
  C+DF^{-1}D^\dag&=&\frac{1}{3}\begin{pmatrix}
\sigma-1 &-2\sigma-1&\sigma-1 \\
\sigma-1&\sigma-1& -2\sigma-1\\
-2\sigma-1&\sigma-1&\sigma-1
\end{pmatrix},
\end{eqnarray}
which has the eigenvalues $\lambda_1=-1$, $\lambda_{2,3}=\mu_{2,3}-1$ and corresponding eigenvectors $v_{i}$, $i=1,2,3$. The optimal scattering matrix $\sigma S_*$ also shares the same eigenvectors, with eigenvalues $\zeta_1=-\sigma$ and $\zeta_{2,3}=\lambda_{2,3}$. Therefore, for a perfectly matched unitary scatterer with $\sigma=1$, the optimal state is reached exactly: $C+DF^{-1}D^\dag=S_*$. This shows that the linear scattering matrix derived in this section successfully approximates the dominant sub-spectrum of $\sigma S_*$.

Using the above intermediate result and the Moore-Penrose Pseudoinverse\footnote{Pedergnana, T. \& Noiray, N. Superradiant scattering by a limit cycle. \textit{Phys. Rev. Appl.} \textbf{20}, 034068 (2023). \url{https://doi.org/10.1103/PhysRevApplied. 20.034068}.}, the non-linear scattering matrix is found to be
 \begin{eqnarray}
    S&=& \dfrac{\vert {s}_\mathrm{out}\rangle \langle s_\mathrm{in}\vert}{\langle s_\mathrm{in}\vert s_\mathrm{in}\rangle} \label{general scattering matrix}\\
    &=&C+\frac{1}{s}D\sum^{3}_{j=1}\begin{pmatrix}
A_{j,+} e^{i \varphi_{j,+}} \\
A_{j,-} e^{i \varphi_{j,-}}
\end{pmatrix} \langle j\vert, \label{Explicit scattering matrix}
\end{eqnarray}
where $A_{j,+}$ and $\varphi_{j,+}$, $j=1,2,3$, are obtained by solving the forced response equations \eqref{forced response A} and \eqref{forced response phi} for $j=1,2,3$.

At large amplitudes $A_\pm >> A_{0,\pm}$, the non-reciprocal amplitude transmission coefficients $T_{j\shortrightarrow i}=\vert (C+DF^{-1}D^\dag)_{ij}\vert$ converge to
\begin{eqnarray}
    \hspace{-0.31cm}&&T_{1\shortrightarrow 2}\approx\bigg\vert\dfrac{2(\sigma^2 + \sigma + 1)}{3(\sigma + 2)}\bigg( \dfrac{e^{-i 2\pi/3}}{1+i(\omega-\omega_+)/\gamma_+}+\dfrac{e^{i 2\pi/3}}{1+i(\omega-\omega_-)/\gamma_-} \bigg)\bigg\vert, \label{Trans12} \\
    \hspace{-0.31cm}&&T_{1\shortrightarrow 3}\approx\bigg\vert\dfrac{2(\sigma^2 + \sigma + 1)}{3(\sigma + 2)}\bigg( \dfrac{e^{i 2\pi/3}}{1+i(\omega-\omega_+)/\gamma_+}+\dfrac{e^{-i 2\pi/3}}{1+i(\omega-\omega_-)/\gamma_-} \bigg)\bigg\vert. \label{Trans13}
\end{eqnarray}
Perfect matching is achieved for $\omega_\pm=\omega\pm\gamma_\pm\sqrt{3}\sigma/(\sigma+2)$, leading to $T_{1\shortrightarrow 3}=(2\sigma+1)/3$ and $T_{1\shortrightarrow 2}=\vert 1-\sigma\vert/3$. 

\subsection*{Large amplitude limit}
Equations \eqref{Trans12} and \eqref{Trans13} describe the asymptotically linear scattering process which occurs in the limit of infinitely large incident wave amplitudes. To understand how the non-linear terms in the modal dynamics \eqref{modal dynamics 2} affect the transmission coefficients in the large (but not infinite) amplitude limit, we use a perturbation series approach. For this, we introduce the small, auxiliary parameter  $\varepsilon > 0$. In the following, we assume forcing from port  ``1'': $\vert s_\mathrm{in}\rangle=s e^{i \omega t}\vert 1 \rangle$, where the incident wave amplitude $s$ is assumed to be of order $\varepsilon^{-1}$, i.e., large. For $\vert a \vert \rightarrow \infty$, the coupled-mode equations \eqref{modal dynamics 2} and \eqref{input-output 2} reduce to Eqs. \eqref{modal dynamics 2, large A} and \eqref{input-output 2, large A}. These limiting cases predict that both $a$ and $\vert s_\mathrm{out}\rangle$ depend linearly on $\vert s_\mathrm{in}\rangle$ in the large amplitude limit. Therefore, we assume that the leading order terms in $a$ and $\vert s_\mathrm{out}\rangle$ scale also with $s$, or $1/\varepsilon$, in the large amplitude limit. We later set $\varepsilon$ equal to $1$ and verify analytically that the assumed scaling ($a\propto s$) is indeed reproduced by our perturbation approach, applied to the non-linear coupled-mode theory described the previous section. To compute the perturbation series, we define the rescaled quantities
\begin{eqnarray}
    \tilde{a}_\pm&=& \varepsilon a_\pm, \label{rescaled amplitude}\\
        \vert \tilde{s}_\mathrm{in}\rangle&=& \varepsilon \vert {s}_\mathrm{in}\rangle, \label{rescaled incident wave}\\
          \vert \tilde{s}_\mathrm{out}\rangle&=& \varepsilon \vert {s}_\mathrm{out}\rangle, \label{rescaled outgoing wave}
\end{eqnarray}
which, by the above assumptions, are all of order $O(1)$. Substituting Eqs. \eqref{rescaled amplitude}-\eqref{rescaled outgoing wave} into the non-linear coupled-mode theory \eqref{modal dynamics 2} and \eqref{input-output 2} yields, after cancelling out $\varepsilon$,
\begin{eqnarray}
    {\dot{\tilde{a}}_{\pm}}&=&{i\omega_{\pm} \tilde{a}_{\pm}} -{\gamma_\pm \tilde{a}_{\pm}}+\dfrac{\varepsilon^2\beta_\pm \tilde{a}_\pm }{\varepsilon^2+{\kappa_\pm \vert \tilde{a}_\pm\vert^2}}  +{D^\dag_\pm \vert \tilde{s}_{\mathrm{in}}\rangle},\label{expanded modal dynamics 2}\\
  \vert \tilde{s}_\mathrm{out}\rangle&=&C \vert \tilde{s}_{\mathrm{in}}\rangle + D \tilde{a}. \label{expanded input-output 2}
\end{eqnarray}
Since $\varepsilon$ was assumed to be small, the rational function in Eq. \eqref{expanded modal dynamics 2} can be expanded as a geometric series:
\begin{eqnarray}
    \dfrac{\varepsilon^2\beta_\pm \tilde{a}_\pm }{\varepsilon^2+{\kappa_\pm \vert \tilde{a}_\pm\vert^2}} &=&\dfrac{\varepsilon^2\beta_\pm \tilde{a}_\pm}{\kappa_\pm \vert \tilde{a}_\pm\vert^2}\Big( 1-\frac{\varepsilon^2}{\kappa_\pm \vert \tilde{a}_\pm\vert^2}+\dots \Big). \label{geo series}
\end{eqnarray}
Next, we expand $\tilde{a}_\pm$ and the outgoing wave in terms of the perturbation parameter:
\begin{eqnarray}
    \tilde{a}_\pm&=& \tilde{a}_{\pm,0}+\varepsilon^2 \tilde{a}_{\pm,1}+O(\varepsilon^4), \label{expansion amplitude}\\
     \vert {s}_\mathrm{out}\rangle&=&  \vert \tilde{s}_\mathrm{out,0}\rangle+\varepsilon^2  \vert \tilde{s}_\mathrm{out,1}\rangle+O(\varepsilon^4). \label{expansion outgoing wave}
\end{eqnarray}
Using the general expression \eqref{general scattering matrix}, the expansion \eqref{expansion outgoing wave} also implies that the scattering matrix can be expanded analogously:
\begin{eqnarray}
  S &=& \dfrac{\big[\vert \tilde{s}_\mathrm{out,0}\rangle+\varepsilon^2  \vert \tilde{s}_\mathrm{out,1}\rangle+O(\varepsilon^4)\big]\langle s_\mathrm{in}\vert}{\langle s_\mathrm{in}\vert s_\mathrm{in}\rangle} \nonumber\\
  &=&S_0+\varepsilon^2  S_1+O(\varepsilon^4)\,. \label{expansion S matrix}
\end{eqnarray}
Because the modal dynamics \eqref{modal dynamics 2} contain a quadratic non-linearity, only even powers of $\varepsilon$ appear in the geometric series \eqref{geo series}. Accordingly, only even orders of $\varepsilon$ contribute to the expansions \eqref{expansion amplitude} and \eqref{expansion outgoing wave}. 

By substituting Eqs. \eqref{geo series}-\eqref{expansion outgoing wave} into the rescaled system \eqref{expanded modal dynamics 2} and \eqref{expanded input-output 2}, we obtain the contributions of different orders in $\varepsilon$ to the scattering matrix. To zeroth order $\varepsilon=0$, as expected, this procedure yields the modal equations in the limit of infinitely large amplitudes:
\begin{eqnarray}
    {\dot{\tilde{a}}_{\pm,0}}&=&{i\omega_{\pm} \tilde{a}_{\pm,0}} -{\gamma_\pm \tilde{a}_{\pm,0}}+{D^\dag_\pm \vert \tilde{s}_{\mathrm{in}}\rangle},\label{expanded modal dynamics 3}\\
  \vert \tilde{s}_\mathrm{out,0}\rangle&=&C \vert \tilde{s}_{\mathrm{in}}\rangle + D \tilde{a}_{0}. \label{expanded input-output 3}
\end{eqnarray}
As standard in the TCMT literature, we assume synchronized conditions in the steady state, $\dot{\tilde{a}}_\pm=i\omega \tilde{a}_\pm$, which leads to 
\begin{eqnarray}
    \tilde{a}_{\pm,0}=\dfrac{D^\dag_\pm \vert \tilde{s}_\mathrm{in}\rangle}{i(\omega-\omega_\pm)+\gamma_\pm}. \label{Solution a0}
\end{eqnarray}
Therefore, the leading-order contribution to the scattering matrix at large amplitudes is the TCMT scattering matrix 
\begin{eqnarray}
    S_0=C+DF^{-1}D^\dag,
\end{eqnarray} 
where the matrix $F$ is defined in Eq. \eqref{Definition F}. Following the above steps and using the expansion \eqref{geo series} we obtain, to order $O(\varepsilon^2)$,
\begin{eqnarray}
    {\dot{\tilde{a}}_{\pm,1}}&=&{i\omega_{\pm} \tilde{a}_{\pm,1}} -{\gamma_\pm \tilde{a}_{\pm,1}}+\dfrac{\beta_\pm \tilde{a}_{\pm,0} }{{\kappa_\pm \vert \tilde{a}_{\pm,0}\vert^2}} ,\label{expanded modal dynamics 4}\\
  \vert \tilde{s}_\mathrm{out,1}\rangle&=&D \tilde{a}_{1}. \label{expanded input-output 4}
\end{eqnarray}
Substituting Eq. \eqref{Solution a0} into Eq. \eqref{expanded input-output 4} and solving for $\tilde{a}_{\pm,1}$ yields
\begin{eqnarray}
    \tilde{a}_{\pm,1}=\dfrac{(\omega-\omega_\pm)^2+\gamma_\pm^2}{\kappa_\pm \langle \tilde{s}_\mathrm{in}\vert D_\pm D^\dag_\pm \vert \tilde{s}_\mathrm{in}\rangle}\dfrac{\beta_\pm D^\dag_\pm \vert \tilde{s}_\mathrm{in}\rangle}{[i(\omega-\omega_\pm)+\gamma_\pm]^2}. \label{Solution a1}
\end{eqnarray}
Using Eq. \eqref{expanded input-output 4} and setting $\varepsilon=1$, then, the next-to-leading order term in the scattering matrix expansion \eqref{expansion S matrix} is found to be
\begin{eqnarray}
    S_1=DF_1^{-1}(s) D^\dag, \label{order 1 S contribution}
\end{eqnarray}
where $s=\sqrt{\langle{s}_\mathrm{in}\vert {s}_\mathrm{in}\rangle}$ is the incident wave amplitude and
\begin{eqnarray}
    F_1^{-1}(s)&=&\begin{pmatrix}
   \theta_+(s) & 0\\
    0 & \theta_-(s)
    \end{pmatrix}, \label{Definition F1}\\
    \theta_\pm(s)&=&\dfrac{(\omega-\omega_\pm)^2+\gamma_\pm^2}{\kappa_\pm \langle {s}_\mathrm{in}\vert D_\pm D^\dag_\pm \vert {s}_\mathrm{in}\rangle}\dfrac{\beta_\pm}{[i(\omega-\omega_\pm)+\gamma_\pm]^2}. 
\end{eqnarray}
As assumed above in the expansion \eqref{expansion S matrix}, $S_1$ is of order $O(\varepsilon^2)=O(s^{-2})$. For infinitely large incident wave amplitudes $s\rightarrow \infty$, $S_1$ and all higher expansion terms vanish. One could continue the perturbation series to show that the next-to-next-to-leading order term is of order $O(s^{-4})$ etc. Therefore, for large $s$, the term \eqref{order 1 S contribution} will dominate the higher-order terms in the expansion \eqref{expansion S matrix}, justifying our perturbative approach. Accordingly, all higher-order contributions to the scattering matrix in the large amplitude limit are neglected in the following.

Collecting the above results, we obtain the first-order correction to the scattering matrix due to the non-linearity in the modal dynamics \eqref{modal dynamics 2}:
\begin{eqnarray}
    S= C+D \big[F^{-1}+F_1^{-1}(s) \big]D^\dag + O(s^{-4}).
\end{eqnarray}
To obtain a simple analytical expression for the scattering matrix at optimally tuned conditions, we assume perfectly matched frequencies: $\omega_\pm=\omega\pm\gamma_\pm\sqrt{3}\sigma/(\sigma+2)$. As in the numerical examples shown in the main text, we set $\kappa_\pm=\kappa$. These simplifications lead to
\begin{eqnarray}
  S_{21}&=& \dfrac{\sigma-1}{3} -\dfrac{(3-6\sigma-6\sigma^2)(\beta_++\beta_-)- 3\sqrt{3}i(1+2\sigma)(\beta_- - \beta_+)}{6 \kappa s^2 (\sigma^2 + \sigma + 1)} \nonumber \\ &&+O(s^{-4}), \label{T12lim} \\
    S_{31}&=&-\dfrac{ (2\sigma+1)  }{3}-\dfrac{(3+12\sigma+3\sigma^2)(\beta_++\beta_-) + 3\sqrt{3}i(1-\sigma^2)(\beta_- - \beta_+) }{6 \kappa s^2 (\sigma^2 + \sigma + 1)}\nonumber\\ &&+O(s^{-4}). \label{T13lim}
\end{eqnarray}
We can now rewrite the approximate amplitude transmission coefficients as follows:
\begin{eqnarray}
 T_{1\shortrightarrow 2}&=& \sqrt{\bigg[\dfrac{1-\sigma}{3} -\dfrac{(6\sigma^2+6\sigma-3)(\beta_++\beta_-)}{6 \kappa s^2 (\sigma^2 + \sigma + 1)}\bigg]^2+\bigg[\dfrac{3\sqrt{3}(1+2\sigma)(\beta_- - \beta_+)}{6 \kappa s^2 (\sigma^2 + \sigma + 1)}\bigg]^2} \nonumber\\ &&+O(s^{-4}), \label{T12lim3} \\
   T_{1\shortrightarrow 3}&=&\sqrt{\bigg[\dfrac{ (2\sigma+1)  }{3}+\dfrac{(3+12\sigma+3\sigma^2)(\beta_++\beta_-)}{6 \kappa s^2 (\sigma^2 + \sigma + 1)}\bigg]^2 + \bigg[\dfrac{3\sqrt{3} (1-\sigma^2)(\beta_- - \beta_+) }{6 \kappa s^2 (\sigma^2 + \sigma + 1)}\bigg]^2}\nonumber\\ &&+O(s^{-4}). \label{T13lim3}
\end{eqnarray}
In the following, we assume that the second brackets under the square roots in Eqs. \eqref{T12lim3} and \eqref{T13lim3}, respectively, is small, which is generally true for small asymmetry $\beta_+ \approx \beta_-$. This leads to
\begin{eqnarray}
 T_{1\shortrightarrow 2}&\approx& \bigg \vert \dfrac{1-\sigma}{3} -\dfrac{(2\sigma^2+2\sigma-1)(\beta_++\beta_-)}{2 \kappa s^2 (\sigma^2 + \sigma + 1)}\bigg \vert+O(s^{-4}),\nonumber\\ \label{T12lim4} \\
   T_{1\shortrightarrow 3}&\approx&\bigg \vert \dfrac{ (2\sigma+1)  }{3}+\dfrac{(1+4\sigma+\sigma^2)(\beta_++\beta_-)}{2 \kappa s^2 (\sigma^2 + \sigma + 1)}\bigg \vert+O(s^{-4}).\nonumber\\ \label{T13lim4}
\end{eqnarray}
We also assume that 
\begin{itemize}
    \item $\sigma>(\sqrt{3}-1)/2\approx 0.366$, so that the term $2\sigma^2+2\sigma-1$ in Eq. \eqref{T12lim4} is positive.
    \item That the $s^{-2}$-modulated term in Eq. \eqref{T12lim4} is less than or equal to $(1-\sigma)/3$, such that the  respective expression inside the vertical lines is non-negative.
\end{itemize}
The first of these assumptions is a minor restriction, since values of $\sigma \in [0.4, 1]$ represent a broad range of dissipative conditions (see also the next section for a quantitative discussion of this point). The second of the above assumptions is generally true for large enough $s$. We note that, in the parameter range considered, $T_{1\shortrightarrow 2}$ is decreased and $  T_{1\shortrightarrow 3}$ is increased by the non-linear effects described the modal dynamics \eqref{modal dynamics 2} for $\beta_\pm >0$. For dissipative conditions ($\sigma<1$), the non-linear interference with the limit cycle compensates--at least partially--for the dissipation incurred at port ``3'', while simultaneously decreasing transmission to port ``2''. To summarize, the above results demonstrate analytically that the non-linear wave-mode coupling envisioned in this work enhances the non-reciprocity of the 3-port scatterer at large incident wave amplitudes, while simultaneously decreasing transmission losses.

\subsection*{Unitarity and reversiblity}
We model dissipation, or irreversible losses, in the infinite amplitude limit \eqref{modal dynamics 2, large A} and \eqref{input-output 2, large A} using the intrinsic decay rates $\gamma_{\mathrm{i},\pm}\in\mathbb{R}^+$. Specifically, we assume that the \cc{decay} rates $\gamma_\pm$ can be decomposed into a part $\gamma_{\mathrm{r},\pm}$ related to (reversible) decay through the ports \cc{(radiation losses)} and another part $\gamma_{\mathrm{i},\pm}$ representing irreversible losses:
\begin{eqnarray}
     \gamma_\pm=\gamma_{\mathrm{r},\pm}+\gamma_{\mathrm{i},\pm}.
\end{eqnarray}
The decay through the ports is reversible because the wave energy transmitted through the ports is available to the environment, while the irreversible losses are irrecoverable. In our theory, the presence of intrinsic losses is marked by a broken time-symmetry: while in forward time, the perturbations $a'_\pm$ decay like $\exp{(-\gamma_\pm t)}$, their time-reversed counterparts $(a'_\pm)^*$ grow like $\exp{(\gamma_{\mathrm{r},\pm} t)}$ in backward time, i.e., only the reversible part of the decay rate $\gamma_{\mathrm{r},\pm}$ contributes to the growth of time-reversed perturbations. Following the work of Zhao, Guo and Fan\footnote{Zhao, Z., Guo, C. \& Fan, S. Connection of temporal coupled-mode-theory formalisms for a resonant optical system and its time-reversal conjugate. \textit{Phys. Rev. A} \textbf{99} (3) (2019). \url{https://doi.org/10.1103/PhysRevA.99.033839}}, the balance of reversible energy is expressed as 
\begin{eqnarray}
     D^\dag D&=&2\begin{pmatrix}
\gamma_{\mathrm{r},+}&0 \\
0&\gamma_{\mathrm{r},-}
\end{pmatrix} \nonumber\\
&=&2\begin{pmatrix}
\gamma_+-\gamma_{\mathrm{i},+}&0 \\
0&\gamma_--\gamma_{\mathrm{i},-}
\end{pmatrix}.\label{Definition internal decay rates}
\end{eqnarray}
To confirm the broken time symmetry, we apply a time-reversal transformation to the exponential decay of the perturbations $a'$ in an initially energized cavity, which is governed by Eq. \eqref{modal dynamics 2, large A}, in the absence of forcing $\vert s_\mathrm{in}\rangle=0$. This transformation interchanges $\vert s_\mathrm{in}\rangle$ and $\vert s_\mathrm{out}\rangle$ in Eq. \eqref{modal dynamics 2, large A}. In the time-reversed scenario, time-reversed perturbations $(a')^*$ grow exponentially in backward time with a priori undetermined growth rates $\tilde{\gamma}_\pm\in\mathbb{R}^+$. The time-reversed dynamics are therefore given by
\begin{eqnarray}
\dot{a}'^*(-t)=(i\omega_0 +\tilde{\gamma}_\pm)a'^*(-t) -D_\pm^T \vert s_\mathrm{out}\rangle^*(-t), \label{time-reversed linearized modal dynamics}
\end{eqnarray}
where $(\cdot)^T$ is the transpose. At the complex frequencies $\omega=\omega_0-i\tilde{\gamma}_\pm$, we have 
\begin{eqnarray}
     2\begin{pmatrix}
\tilde{\gamma}_+&0 \\
0&\tilde{\gamma}_-
\end{pmatrix} a'^*&=& D^T \vert s_\mathrm{out}\rangle^*\\
&=& D^T(D^*a'^*)\\
\Rightarrow D^\dag D &=&2\begin{pmatrix}
\tilde{\gamma}_+&0 \\
0&\tilde{\gamma}_-
\end{pmatrix}. \label{auxiliary eq. 1}
\end{eqnarray}
Comparing Eqs. \eqref{auxiliary eq. 1} and \eqref{Definition internal decay rates} yields 
\begin{eqnarray}
     \tilde{\gamma}_\pm=\gamma_\pm-\gamma_{\mathrm{i},\pm}=\gamma_{\mathrm{r},\pm},
\end{eqnarray}
demonstrating that time-symmetry is broken ($\gamma_\pm\neq \tilde{\gamma}_\pm$) for $\gamma_{\mathrm{i},\pm}\neq 0$. Using Eqs. \eqref{coupling matrix D} and \eqref{Definition internal decay rates}, a relation between the intrinsic decay rates $\gamma_{\mathrm{i},\pm}$ and the unitarity factor $\sigma$ is established:
\begin{eqnarray}
     \dfrac{\gamma_{\mathrm{i},\pm}}{\gamma_\pm}=\dfrac{1-\sigma^2}{\sigma+2}\geq0. \label{relation irrev. and sigma}
\end{eqnarray}
The non-unitarity and the irreversible losses of the scattering process are intrinsically connected. In particular, unitary scattering ($\sigma=1$) corresponds to perfect reversibility ($\gamma_{\pm}=\gamma_{\mathrm{r},\pm}$). It was mentioned in the previous section that $1>\sigma>0.4$ describes a broad range of dissipative conditions. We see from Eq. \eqref{relation irrev. and sigma} that, at $\sigma=0.4$, the irreversible decay rates $\gamma_\mathrm{i,\pm}=0.35 \gamma_\pm$ make up more than a third of the total decay rates $\gamma_\pm$ in the limit of infinitely large amplitudes.




\end{document}